%% file: main.tex
\documentclass[sigplan,screen]{acmart}

\copyrightyear{2025}
\acmYear{2025}
\setcopyright{acmlicensed}
\acmConference[ASPLOS '25] {Proceedings of the 29th ACM International Conference on Architectural Support for Programming Languages and Operating Systems, Volume 1}{March 30--April 3, 2025}{Rotterdam, Netherlands.}
\acmBooktitle{Proceedings of the 29th ACM International Conference on Architectural Support for Programming Languages and Operating Systems, Volume 1 (ASPLOS '25), March 30--April 3, 2025, Rotterdam, Netherlands}
\acmISBN{979-8-4007-0698-1/25/03}
\acmDOI{10.1145/XXXXXX.XXXXXX}

\usepackage[]{hyperref}

\author{Haiyu Huang}
\email{huanghy95@mail2.sysu.edu.cn}
\orcid{0009-0000-6146-2493}
\affiliation{%
  \institution{Sun Yat-sen University}
  \city{Guangzhou}
  \country{China}
}

\author{Cheng Chen}
\email{wu.cc@alibaba-inc.com}
\affiliation{%
  \institution{Alibaba Group}
  \city{Hangzhou}
  \country{China}
}

\author{Kunyi Chen}
\email{chenkunyi.cky@alibaba-inc.com}
\affiliation{%
  \institution{Alibaba Group}
  \city{Hangzhou}
  \country{China}
}

\author{Pengfei Chen*}
\orcid{0000-0003-0972-6900}
\email{chenpf7@mail.sysu.edu.cn}
\affiliation{%
  \institution{Sun Yat-sen University}
  \city{Guangzhou}
  \country{China}
}

\author{Guangba Yu}
\email{yugb5@mail2.sysu.edu.cn}
\orcid{0000-0001-6195-9088}
\affiliation{%
  \institution{Sun Yat-sen University}
  \city{Guangzhou}
  \country{China}
}

\author{Zilong He}
\orcid{0000-0001-7963-082X}
\email{hezlong@mail2.sysu.edu.cn}
\affiliation{%
  \institution{Sun Yat-sen University}
  \city{Guangzhou}
  \country{China}
}

\author{Yilun Wang}
\email{wangylun6@mail2.sysu.edu.cn}
\orcid{0000-0001-9262-8652}
\affiliation{%
  \institution{Sun Yat-sen University}
  \city{Guangzhou}
  \country{China}
}

\author{Huxing Zhang}
\email{huxing.zhx@alibaba-inc.com}
\affiliation{%
  \institution{Alibaba Group}
  \city{Hangzhou}
  \country{China}
}

\author{Qi Zhou}
\email{jackson.zhouq@alibaba-inc.com}
\affiliation{%
  \institution{Alibaba Group}
  \city{Hangzhou}
  \country{China}
}

\thanks{
* Pengfei Chen is the corresponding author.}

\renewcommand{\shortauthors}{Huang et al.}

\usepackage{tcolorbox}
\usepackage{multirow}
\usepackage{threeparttable}
\usepackage{colortbl}

\begin{document}

\title[Mint: Cost-Efficient Tracing via Commonality and Variability Analysis]{Mint: Cost-Efficient Tracing with All Requests Collection via Commonality and Variability Analysis}


\begin{abstract}
Distributed traces contain valuable information but are often massive in volume, posing a core challenge in tracing framework design: balancing the tradeoff between preserving essential trace information and reducing trace volume. To address this tradeoff, previous approaches typically used a `1 or 0' sampling strategy: retaining sampled traces while completely discarding unsampled ones. However, based on an empirical study on real-world production traces, we discover that the `1 or 0' strategy actually fails to effectively balance this tradeoff. 

To achieve a more balanced outcome, we shift the strategy from the `1 or 0' paradigm to the `commonality + variability' paradigm. The core of `commonality + variability' paradigm is to first parse traces into common patterns and variable parameters, then aggregate the patterns and filter the parameters. We propose a cost-efficient tracing framework, \textit{Mint}, which implements the `commonality + variability' paradigm on the agent side to enable all requests capturing. Our experiments show that \textit{Mint} can capture all traces and retain more trace information while optimizing trace storage (reduced to an average of 2.7\%) and network overhead (reduced to an average of 4.2\%). Moreover, experiments also demonstrate that \textit{Mint} is lightweight enough for production use. 

\end{abstract}

\maketitle 

\section{Introduction}
As software systems grow larger and more complex~\cite{yu2023logreducer}, distributed tracing has become a critical infrastructure, providing visibility into systems' end-to-end runtime behavior~\cite{sifter}. Tracing frameworks like Jaeger~\cite{jaeger}, OpenTelemetry~\cite{opentelemetry}, and Zipkin~\cite{zipkin} have been widely adopted by major internet companies~\cite{hindsight}. The trace data they generate, which visualizes the end-to-end paths of requests through service instances, has proven to be extremely helpful for profiling 
systems \cite{GMTA}, detecting anomalies~\cite{tracead, traceCRL, microhecl}, and diagnosing failures~\cite{TraceAnomaly, gan2023sleuth, Microrank, Sage}.

\begin{figure}[t]
\centering
\includegraphics[width=0.88\linewidth]{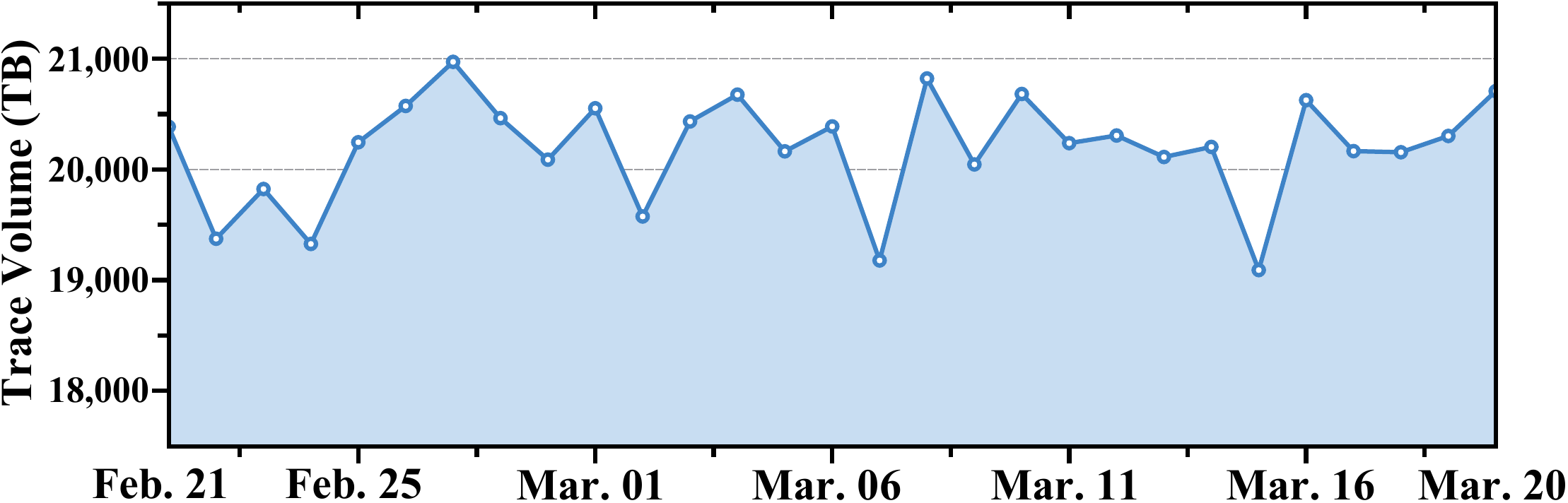}
\caption{A tracing system in Alibaba generates 18.6-20.5 PBs of traces per day between Feb. 21 and Mar. 20, 2024.}
\label{fig:trace-volume}
\end{figure}

Although distributed traces are helpful, they are often voluminous~\cite{hindsight}, making their collecting,  storing, and processing extremely expensive, especially in production environments~\cite{sifter}. For instance, as shown in Fig.~\ref{fig:trace-volume}, a large-scale e-commerce system in Alibaba \cite{alibaba} generates approximately 18.6-20.5 pebibytes (PBs) of traces per day. Therefore, reducing tracing overhead and efficiently preserving the valuable information within trace data at an acceptable cost is a crucial and significant task~\cite{wset}.

In current tracing systems, the de facto practice to handle this task is through trace sampling (i.e., retaining only a portion of traces)~\cite{sigelman2010dapper, kaldor2017canopy, wset, sifter, samplehst, sieve, hindsight, trastrainer}. The main workflow is to first determine which traces should be sampled, and then retain 
the sampled traces while completely discarding the unsampled ones (we call it the `1 or 0' strategy). Depending on the timing of the sampling decision and the sampling rules, these methods are typically categorized as head sampling~\cite{sigelman2010dapper, kaldor2017canopy}, tail sampling~\cite{sifter, wset, samplehst, sieve}, and the recently introduced retroactive sampling~\cite{hindsight}.


However, our research revealed significant shortcomings of the prevailing trace sampling techniques utilising the `1 or 0' strategy, as evidenced by an empirical trace study (\S~\ref{subsec:empi}) conducted on real-world systems.
(1) \textbf{The drawback of completely discarding unsampled traces.}
While current sampling methods try to retain valuable traces through certain rules, our findings indicate that those discarded traces may also be queried by Site reliability engineers (SREs) because the characteristics of the traces needing analysis are often unpredictable. This is evidenced by the observation that the current sampling strategy results in a query miss rate of approximately 27.17\% in our study. Trace query failure is a potential source of impediment to the SRE diagnosis process. 
(2) \textbf{Lack of effective compression of individual trace volumes.} Previous trace reduction methods only reduce the number of traces without lightweighting each individual trace. However, each trace can contain more detailed information than debug-level log~\cite{sigelman2010dapper}, making it necessary to compress traces based on their characteristics. On the other hand, general-purpose compression tools~\cite{bzip2, gzip, lzma} and previously proposed log compression techniques~\cite{liu2019logzip, liu2019logzip, CLP} are ineffective for trace compression because traces have a topological data structure. These methods fail to fully utilize trace characteristics, resulting in poor compression performance (details shown in \S~\ref{subsubsec:expr_rq3}).

To address the above limitations,  we shift the strategy of trace overhead reduction from the `1 or 0' paradigm to the \textbf{`commonality + variability'} paradigm which parses trace data into common patterns and variable parameters, and processes them individually. Through our empirical study (\S~\ref{subsubsec:empi_rq3}), we find that the widely existing commonality and variability in traces can be leveraged to preserve more trace information at a lower cost. By leveraging commonality (e.g., constructing common patterns), we can cluster and store the basic information of all traces at a low cost. Moreover, by utilizing variability (e.g., extracting parameters), we can better filter and efficiently record the differing parts.

To implement this, we develop \textit{\textbf{Mint}}, a cost-efficient distributed tracing framework that captures all requests and retains near-full trace information. The workflow of \textit{Mint} is to first analyze commonality and variability at two levels within traces to parse them into patterns and parameters.
Then it mounts metadata of all traces onto their corresponding patterns using a low-cost method (i.e., Bloom Filter~\cite{bloom}), and filters and retains valuable information from the variable parameters. \textit{Mint} does not directly discard the unsampled traces, under the `commonality + variability' paradigm, the difference between handling sampled and unsampled traces is whether the variability part is sent to the tracing backend. For unsampled traces, only the basic information (i.e., commonality part) of them is preserved at a low cost, which is sufficient for analysis. For sampled traces, their full information (i.e., both of commonality and variability part) is retained, and their volume is reduced by compression. As a practical tool, \textit{Mint} reduces traces on the agent generation side, thus saving both network bandwidth and storage space. 

We conducted extensive experiments to verify \textit{Mint}'s effectiveness and performance. 
Experiments show that \textit{Mint} reduces total trace storage overhead to 2.7\% and network overhead to 4.2\%, while recording all requests.
For practical application, \textit{Mint} has been deployed in the production environment of Alibaba for over two months, successfully reducing trace volume while capturing all requests.


In summary, our study makes the following contributions.
\begin{itemize}
    \item We conduct an empirical study on traces in real-world systems and obtain three observations that can facilitate the trace reduction task.
    \item We point out the limitations of current trace reduction methods based on the `1 or 0' paradigm, and introduce the `commonality + variability' paradigm to retain more trace information at a lower cost.
    \item We propose a practical distributed tracing framework named \textit{Mint}, which applies the `commonality + variability' paradigm on the agent side, enabling cost-efficient retention of all requests.
    \item We conduct extensive experiments  to evaluate \textit{Mint} and demonstrate its effectiveness in reducing trace volume while capturing all requests. We also assess \textit{Mint}'s efficiency, showing that it is a practical tool.
\end{itemize}

\section{Background and Motivation}
\subsection{Related Work}
\textbf{Distributed tracing.} Distributed tracing is crucial for providing observability and maintaining highly dynamic microservice systems. Magpie~\cite{magpie}, X-trace~\cite{x-trace}, Dapper~\cite{sigelman2010dapper}, Pinpoint~\cite{pinpoint}, and Pivot~\cite{pivot} are classic distributed tracing frameworks that have introduced essential end-to-end observability into distributed systems. OpenTelemetry~\cite{opentelemetry} offers a unified, high-performance tracing API and SDK, and has established the OTLP~\cite{otlp} standard to standardize the format of trace data and its transmission between services. Jaeger~\cite{jaeger} and Zipkin~\cite{zipkin} are also popular open-source tracing frameworks widely adopted in the industry. Utilizing end-to-end observability from traces, previous works have performed trace analysis in areas such as performance profiling~\cite{mi2013toward, Diagnosing, GMTA}, anomaly detection\cite{tracead, traceCRL, microhecl}, and failure diagnosis~\cite{TraceAnomaly, Microrank, Sage, gan2023sleuth}. 
As trace volume increases, previous studies have also proposed various methods for trace reduction~\cite{milliScope, kaldor2017canopy, sifter}, with mainstream approaches including head sampling~\cite{sigelman2010dapper, kaldor2017canopy}, tail sampling~\cite{sifter, wset, samplehst, sieve}, and retroactive sampling~\cite{hindsight}.

\textbf{Log-specific compressor.} Log data is the cousin of distributed traces, but lacks topology structure like traces. Due to the high redundancy of logs, many methods have been proposed to compress logs. LogArchive~\cite{logarchive}, Cowic~\cite{cowic}, and MLC~\cite{mlc} compress logs by extracting features from log data. Some log compression methods~\cite{logreducer21fast, improving, logzip, elise,loggrep} use a log parser to separate logs and process headers and variables independently for further compression. LogZip~\cite{logzip} and RoughLogs~\cite{Meinig2019RoughLA} compress logs by building a model to identify redundancies. CLP~\cite{CLP} parses logs into schemas, storing variables as dictionary and non-dictionary.
LogGrep~\cite{loggrep} structures and organizes log data into fine-grained units by exploiting both static and runtime patterns. 
However, due to the significant differences in format and structure between distributed traces and logs~\cite{sigelman2010dapper}, directly applying log compression methods to traces does not yield ideal results (as shown in our experimental results in \S~\ref{subsubsec:expr_rq3}).


\subsection{Empirical Study on Traces in Industry}
\label{subsec:empi}
Distributed tracing is immensely helpful for maintaining software systems. However,  as shown in Fig.~\ref{fig:trace-volume}, the significant overhead it introduces cannot be overlooked~\cite{hindsight, sifter}. 
To better leverage trace characteristics for improving trace overhead reduction, we conducted an empirical study on industrial traces from real-world systems at Alibaba. Alibaba is a large commercial cloud service provider, and the traces we collected are from systems deployed on over 1 million nodes and involving more than 20,000 microservices, we believe such a scale of samples can effectively represent traces in large-scale industrial microservices systems. Our study aims to answer the following research questions (RQs):

\textit{\textbf{RQ1:}} What kind of overhead do traces introduce?

\textit{\textbf{RQ2:}} Can existing strategies effectively reduce trace overhead while retaining valuable information?

\textit{\textbf{RQ3:}} What characteristics in trace data can we leverage?

\subsubsection{\textit{RQ1: Tracing Overhead. }}
\label{empi_rq1}


To better understand the overhead caused by tracing, we first investigate the lifecycle of traces. The lifecycle of traces consists of four main stages~\cite{datadog, sifter}. \textcircled{\raisebox{-0.9pt}{1}} \textbf{\textit{Init trace}}: the trace life cycle begins when a request is initiated (e.g., when a user submits a form on a website), and this triggers the creation of a new unique trace ID. \textcircled{\raisebox{-0.9pt}{2}} \textbf{\textit{Generate trace data}}: as the request progresses, each time it passes through related applications, those applications equipped with tracing client libraries generate trace data to capture the events triggered (e.g., alerts) and state data (e.g., error codes or delays). 
This information is recorded in key-value format within the core components of the trace, known as `span'~\cite{span}. 

\textcircled{\raisebox{-0.9pt}{3}} \textbf{\textit{Report trace data}}: at each application node, the trace agent intercepts, collects, serializes, and transmits trace data generated on that node to a centralized trace backend via the network~\cite{hindsight}. \textcircled{\raisebox{-0.9pt}{4}} \textbf{\textit{Store trace data}}: after reported to the backend, trace data from different application nodes with the same trace ID is joined together and stored in a persistent storage device for subsequent analysis. From the life cycle, we can see that stage 
\textcircled{\raisebox{-0.9pt}{3}} involves network overheads, and stage \textcircled{\raisebox{-0.9pt}{4}} involves storage overheads.

\begin{figure}[t]
\centering
\includegraphics[width=\linewidth]{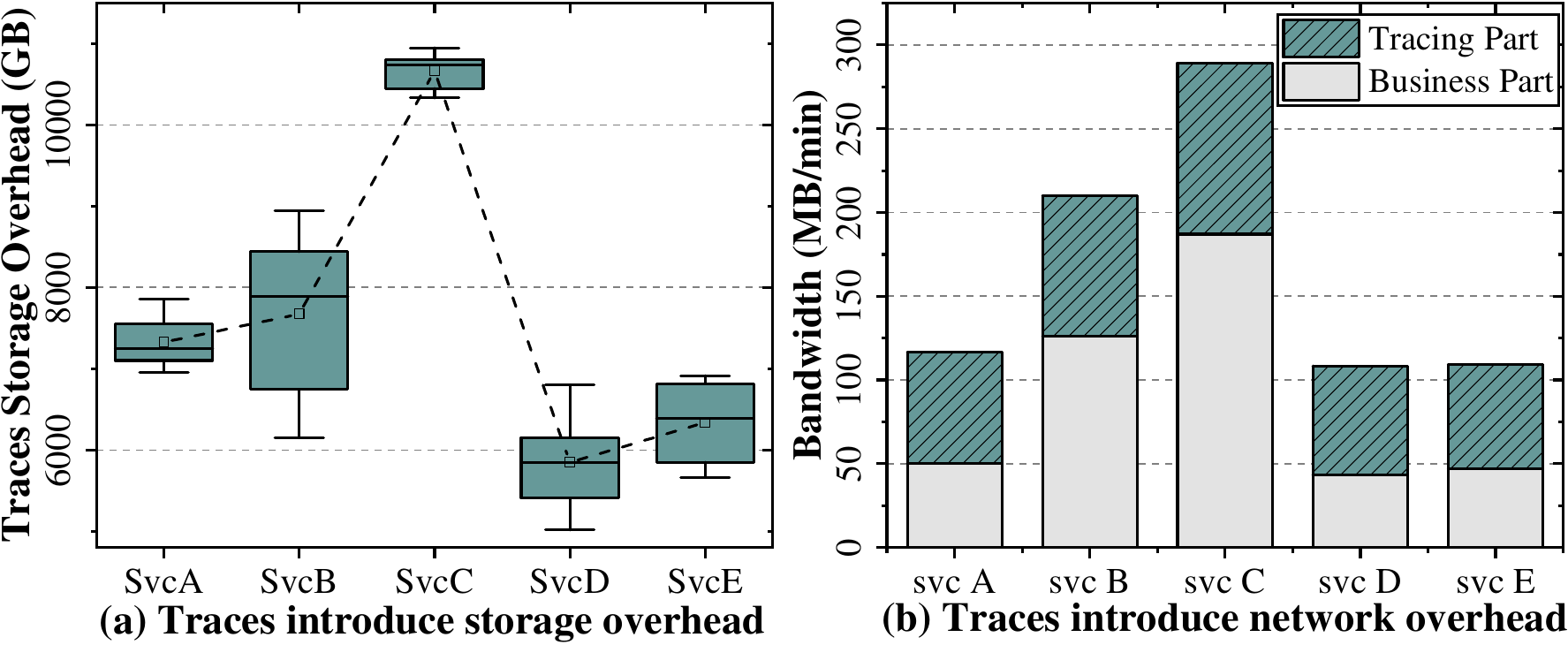}
\caption{Storage overhead and bandwidth increment caused by tracing in 5 services from Alibaba.}
\label{fig:empi_rq1}
\end{figure}

To further explore the impact of these overheads, we investigated the top 5 services in Alibaba with the largest trace volume. We collected and tallied the storage overhead of these services' traces generated from February 21 to March 20, 2024, as well as the bandwidth increment caused by reporting traces during this period, as shown in Fig.~\ref{fig:empi_rq1}.

Fig.~\ref{fig:empi_rq1} (a) shows the storage overhead caused by traces, revealing that these services spent an average of 7,639 GB on storing traces per day. Considering the storage cost of \$0.50/GiB per month, these services would require an average monthly expenditure of \$114.59k to fully store the trace data. Fig.~\ref{fig:empi_rq1} (b) demonstrates that adopting tracing introduces up to 102 MB/min of additional bandwidth between nodes, significantly affecting latency-sensitive application traffic. It is evident that both storage and network overhead caused by tracing are substantial, which can severely impact the scalability and robustness of software systems. 

\begin{tcolorbox}
\textbf{Finding 1.} Traces introduce both costly storage overhead and unignorable network overhead.
\tcblower
\textbf{Implication 1.} To reduce both overheads, we  implement trace reduction earlier on the agent side.
\end{tcolorbox}



\subsubsection{\textit{RQ2: Limitations of Existing Work.}}
\label{subsubsec:RQ2}
To address the critical trace overhead reduction problem, prior work has made some attempts~\cite{sigelman2010dapper, kaldor2017canopy, wset, sifter, samplehst, sieve, hindsight}. 
However, our investigation reveals that these methods still fail to resolve the trade-off between retaining necessary trace information and reducing overhead to a reasonable level, primarily due to the following two limitations.

\begin{figure}[t]
\centering
\includegraphics[width=\linewidth]{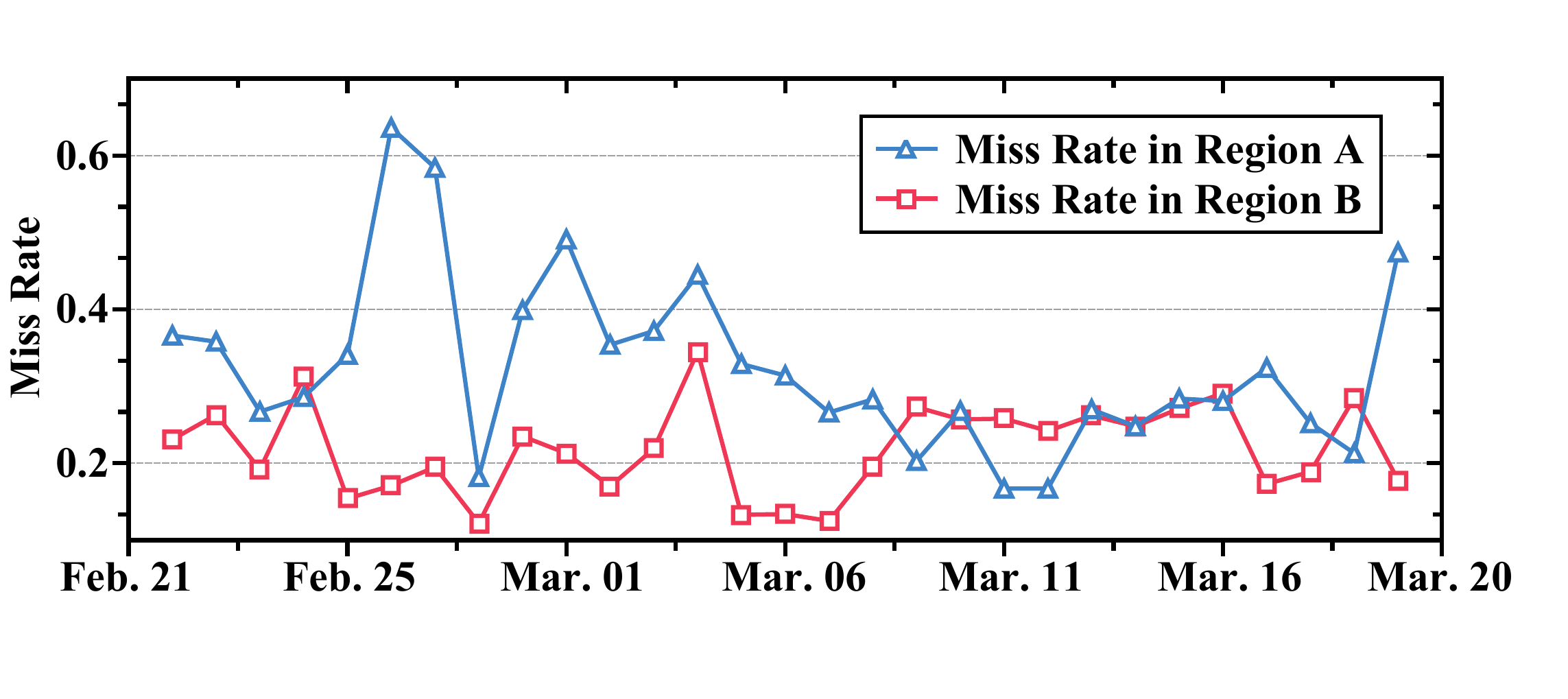}
\caption{The miss rate in two different regions from Alibaba per day between Feb. 21 and Mar. 20, 2024.}
\label{fig:empi_rq2}
\end{figure}

\textbf{The drawback of completely discard traces that are not sampled.} Previous trace reduction methods have been based on a `1 or 0' strategy, meaning they retain the full information of sampled traces while completely discarding traces that are not sampled. However, this strategy can result in the loss of essential information because some traces that need to be analyzed cannot be determined until the analysis is performed. Take a real-world case in Alibaba as an example:
on Mar. 25, 2024, analysts needed to analyze requests that occurred on Mar. 21, with particular \textit{trace\_ids} over a period of time, which cannot be known in advance when generating trace data four days earlier. When analysts query traces that were not sampled but are necessary for analysis, they receive no results. We investigated the proportion of traces that analysts were unable to retrieve due to querying unsampled traces in Alibaba over 30 days, referred to as the miss rate. During this period, Alibaba employed a sampling strategy that combined OpenTelemetry's head sampling~\cite{ot-headsampling} and tail sampling~\cite{ot-tailsampling}. The results, as shown in Fig.~\ref{fig:empi_rq2}, indicate that the average miss rate for trace queries at Alibaba over the past 30 days was 27.17\%, indicating that a significant number of traces that need to be analyzed were filtered out due to sampling. This motivates us to retain essential information from the unsampled traces in a cost-
efficient manner, rather than just discarding them.

\textbf{Lack of effective compression of individual trace volumes.} The `1 or 0' sampling strategy used by previous methods have another limitation: for sampled traces, all their original data is preserved in raw format. This means that prior methods only reduce the number of traces without lightweighting each trace itself~\cite{sigelman2010dapper, kaldor2017canopy, sifter}. However, traces can be detailed and produced at high volume~\cite{hindsight}. At  major internet companies (e.g., Google~\cite{sigelman2010dapper}, Facebook~\cite{facebook}, etc.), traces are typically more detailed than debug-level logging, and each traced request generates several MBs of tracing data~\cite{hindsight}. 
We analyzed the distribution of the volume of trace data generated by each traced request in Alibaba over a day, experiment result shows that more than 11\% of traces exceed 1.2 MB. This motivates us to design compression algorithms tailored to trace data characteristics, aiming not only to reduce the number of traces but also to lightweight the volume of each trace, thus better reducing trace overhead.

\begin{tcolorbox}
\textbf{Finding 2.} Existing trace reduction methods employing a `1 or 0' strategy fail to adequately address the tradeoff between preserving necessary trace information and reducing overhead.
\tcblower
\textbf{Implication 2.} We should design a better trace information retention method tailored to the characteristics of trace data: \textcircled{\raisebox{-0.9pt}{1}} For unsampled traces, retain essential information in a cost-efficient manner rather than discard them entirely. \textcircled{\raisebox{-0.9pt}{2}} For sampled traces, effective compression should be applied instead of storing the full information in raw format.
\end{tcolorbox}

\subsubsection{\textit{RQ3: Trace Data Characteristics.}}
\label{subsubsec:empi_rq3}
\begin{figure}[t]
\centering
\includegraphics[width=0.95\linewidth]{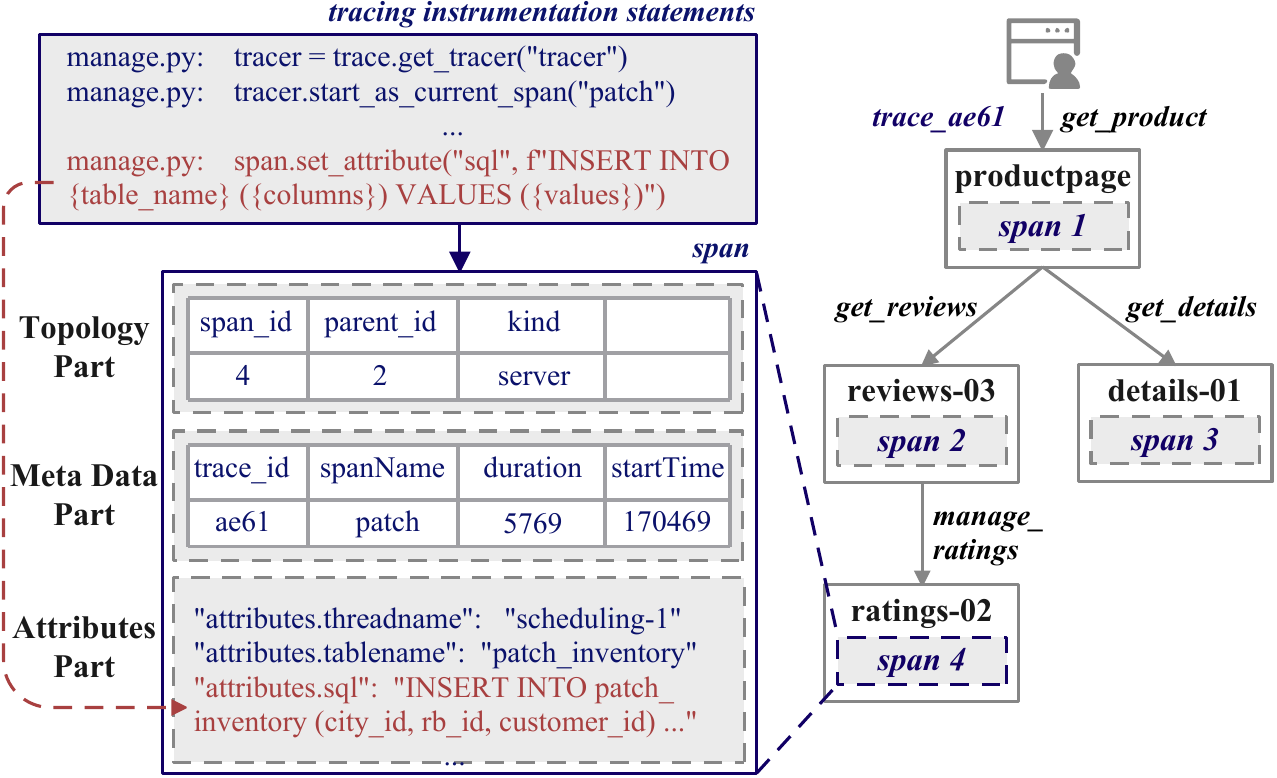}
\caption{An example of trace and span structure.}
\label{fig:trace_structure}
\end{figure}

To investigate the characteristics of traces more effectively, we first need to understand the typical structure of trace data.
As depicted in Fig.~\ref{fig:trace_structure}, each trace forms a tree-like structure composed of a series of spans linked according to their invocation relationships~\cite{tprof}. Each span represents a single unit of work and typically consists of three parts~\cite{span}: (1) Topology part: which includes information indicating the position of the span within the entire trace. 
(2) Metadata part: containing predefined basic information automatically obtained and attached to the span by the client library. (3) Attributes part: where users can include additional detailed information about the invocation process, such as debuggable or identifiable data (e.g., alert logs, SQL query content, etc.). This information is added to the span through attributes or events using tracing instrumentation statements~\cite{instrumentation}, as illustrated in Fig.~\ref{fig:trace_structure}. After investigating 3,419,503 traces generated in Alibaba within a day, we found that commonality and variability are widely present in trace data, and they can be observed in the following levels.

\textbf{Inter-trace level.} Although there are numerous traces, many of them are triggered by the same type of requests~\cite{tprof} (e.g., multiple queries for the same product  information by different users, resulting in traces belonging to the `product query' type). Due to the identical task logic, traces of the same type exhibit strong commonality (i.e., they pass through the same sequence of services in the same order). However, due to different parameters and runtime states, there is variability in the specific information on each span.

\textbf{Inter-span level.} Since a span represents a unit of work, spans that execute the same work logic will have a similar structure~\cite{traceCRL}. For instance, if two spans are generated by the code block shown in Fig.~\ref{fig:trace_structure}'s \textit{manage.py}, they will possess the same keys and their values will also follow a similar pattern. To illustrate, for the `attributes.sql' field, as it originates from the same statement, its value will adhere to the pattern `INSERT INTO xx (xx VALUES xx)'. Variability can be observed in the change of parameters within regular expressions (e.g., `\textit{city\_id}, \textit{rb\_id},  \textit{customer\_id}' of span 4 in Fig.~\ref{fig:trace_structure}).

\input{table/empi_rq3}

To further investigate the frequency of commonality in the two aforementioned levels, we calculated the occurrence and proportion of pairs with commonality (i.e., two traces or spans that have common pattern) to the total number of pairs (i.e., any two different traces or spans) in each level. Table~\ref{tab:empi_rq3} presents the results of our analysis of traces generated within a week across three services. As shown, inter-trace pairs with commonality account for about 34\% - 56\% of all inter-trace pairs, while inter-span pairs with commonality make up around 25\% - 45\% of all inter-span pairs. 

\begin{tcolorbox}
\textbf{Finding 3.} Commonality and variability widely exist among pairs of traces and spans.
\tcblower
\textbf{Implication 3.} We can leverage these commonality and variability to reduce trace overhead through gathering common patterns and efficiently recording the differential parts.
\end{tcolorbox}

\begin{figure*}[t]
\centering
\includegraphics[width=0.85\textwidth]{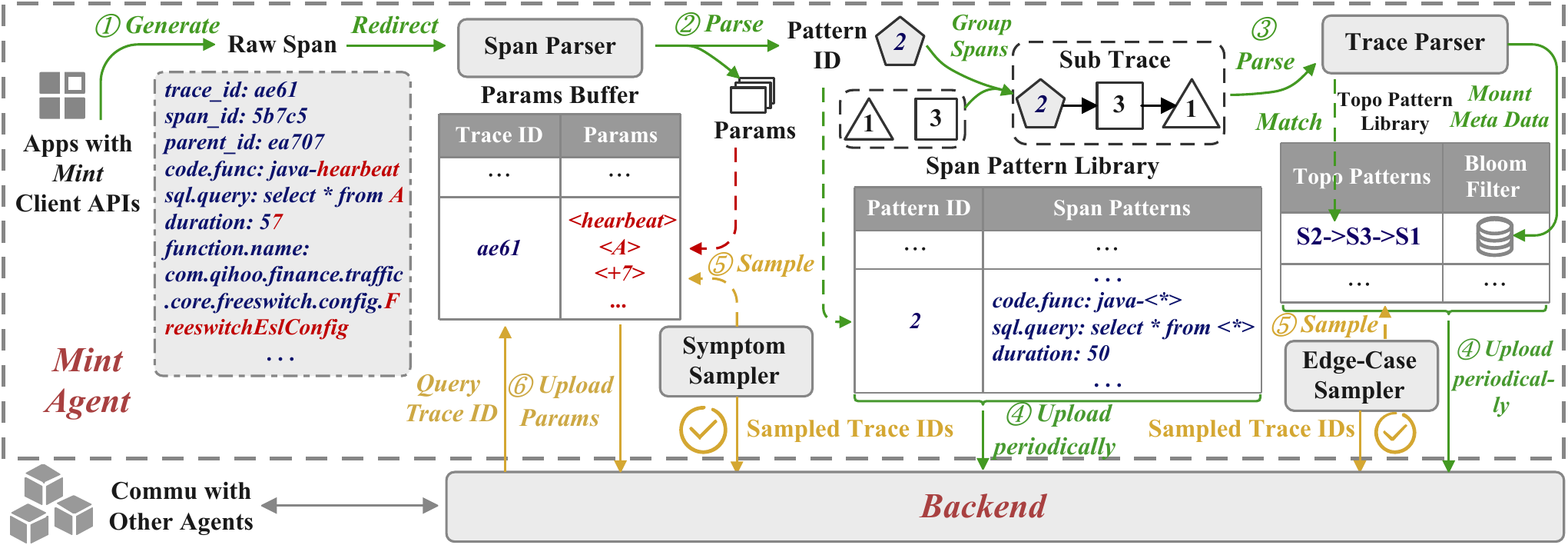}
\caption{An overview of \textit{Mint}'s tracing walkthrough.}
\label{fig:sys_overview}
\end{figure*} 

\section{Techniques}

\subsection{Overview}

Facilitated by the above observations, we propose \textit{Mint}, a tracing framework that aims to tackle the current tradeoff between preserving essential trace information and reducing trace volume by shifting the `1 or 0' paradigm to the `commonality + variability' paradigm on the agent side. Our goal is to allow practitioners to 
capture all requests and retain near-full trace information in a cost-efficient way. 

\textbf{Tracing walkthrough. }We first describe the tracing walkthrough (i.e., the entire process of generating and saving trace data) at a high level using \textit{Mint}, as shown in Fig.~\ref{fig:sys_overview}.

\textcircled{\raisebox{-0.9pt}{1}} \textit{\textbf{Trace Data Generating.}} When a request carrying a trace ID passes through an application node, trace data (i.e., spans) are generated by \textit{Mint}'s client API, similar to existing frameworks. However, instead of immediately recording or reporting the spans, \textit{Mint} redirects them to Span Parser.

\textcircled{\raisebox{-0.9pt}{2}} \textit{\textbf{Inter-Span Level Parsing.}} 
Span Parser analyzes the commonality and variability at span level to parse the incoming span into a pattern and parameters (the blue part and the red part in Fig.~\ref{fig:sys_overview}). 
The pattern updates span Pattern Library and is encoded into a pattern ID, while parameters are temporarily stored in Params Buffer on the agent.

\textcircled{\raisebox{-0.9pt}{3}} \textit{\textbf{Inter-Trace Level Parsing.}} As a single request may traverse different applications on the same node, generating multiple spans, these spans form a tree-like structure based on their call relationships, we call it a sub-trace (i.e., a segment of a trace on the same node). Trace Parser then leverages the commonality and variability between sub-traces to find the most similar pattern in the topology Pattern Library for the incoming sub-trace, marked as the matched template. The metadata (e.g., trace ID) of the coming sub-trace is then mounted on the matched template using a Bloom Filter and stored on the agent.

\textcircled{\raisebox{-0.9pt}{4}} \textit{\textbf{Basic Information Uploading.}} Through the above two parsers, the pattern and basic information of each trace data are cost-effectively stored on the agent side via the Pattern Library and Bloom Filter. \textit{Mint} agent periodically uploads this information to backend to ensure the basic information of all traces is preserved. As for the detailed parameters temporarily stored in the Params Buffer, \textit{Mint} decides whether to emit 
the parameters to backend based on whether the trace they belong to is marked as sampled. 

\textcircled{\raisebox{-0.9pt}{5}} \textit{\textbf{Key Traces Sampling.}} \textit{Mint} employs two samplers: the Symptom Sampler monitors the Params Buffer, marking traces with abnormal values (e.g., status code 502) or outliers (e.g., unusually large duration values) as sampled traces; the Edge-Case Sampler monitors the Pattern Library, marking traces with rare execution paths as sampled traces. 

\textcircled{\raisebox{-0.9pt}{6}} \textit{\textbf{Parameters Uploading.}} Once a trace is marked as sampled on an agent, all parameters of this trace distributed across different agents are emitted to the backend through communication between agents, ensuring trace coherence. 

\subsection{Inter-Span Level Parsing}
Once spans are generated, \textit{Mint} parses them into common patterns and variable parameters, storing them separately. Thus it can reduce tracing overhead by aggregating patterns and selectively retaining only part of parameters. The span parsing process involves the following stages:

\subsubsection{\textit{Offline Stage: Warming up Span Parser.}} 
We first randomly sample $m$ (5,000 in our implementation) raw spans generated on the node over a recent period to build and warm up the parser offline. This offline stage helps achieve acceptable performance in the early stages of online parsing, mitigating cold start issues. Notably, once \textit{Mint} is running stably, it is not sensitive to the early random sampling.

\begin{figure}[t]
\centering
\includegraphics[width=\linewidth]{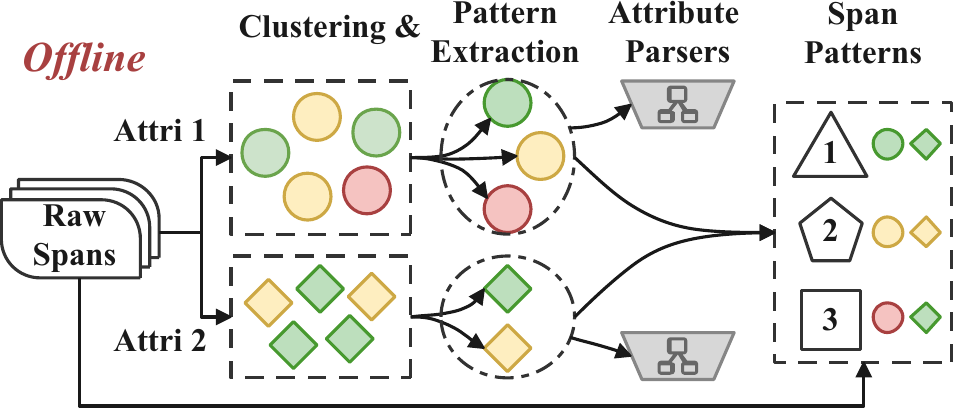}
\caption{The offline stage of span parser.}
\label{fig:offline}
\end{figure}

Fig.~\ref{fig:offline} shows the offline construction process of the span parser. The core idea is to train a parser for each attribute of the span and then combine different attribute patterns to form a complete span pattern. 

\textbf{Clustering and pattern extracting.} Since different attributes have different semantics, to speed up the parsing stage, we train a separate parser for each attribute to avoid meaningless comparisons between different semantics. To implement this, we first 
cluster and extract patterns for each attribute based on its data type:

For attributes with string values, we use the longest common subsequence (LCS) to compute the similarity between string values as follows:
\begin{equation}
\begin{aligned}
\delta (s_1, s_2) = \frac{|LCS(s_1, s_2)|}{\max(|s_1|, |s_2|)},
\end{aligned}
\end{equation}\label{equ:p_s}
where $s_1$ and $s_2$ are tokenized strings (using words as tokens in our implementation), and $| \cdot |$ denotes the number of tokens in a string sequence. For all possible values of the same string-type attribute in sampled spans, we aggregate values with similarity above a threshold (0.8 in our implementation) to form clusters $C=\{C_0, ..., C_n\}$. For each cluster $C_i$, we extract the shortest regular expression that can represent all strings in the cluster, which serves as the pattern $P_i$ for that cluster.

For attributes with numeric values, we use a bucketing approach based on exponential intervals. We first select a precision parameter $\alpha$ (0.5 in our implementation). For each numeric value $d$, we store it in a bucket $B_i$ with index $i=\left\lceil \log_\gamma(d) \right\rceil$, where $\gamma=\frac{1+\alpha}{1-\alpha}$. Thus values in bucket $B_i$ fall within the interval $(\gamma^{i-1}, \gamma^i]$. Specifically, values in bucket $B_0$ fall within $(0, 1]$. This approach clusters numeric values into buckets $B=\{B_0, ..., B_n\}$, with each bucket $B_i$ represented by the interval pattern $(\text{lower}_i, \text{upper}_i]$.

\textbf{Parsers building.} Following the steps above, we extracted a series of patterns $P = \{P_1, ..., P_n\}$ for each attribute $A_i$. \textit{Mint} uses these patterns to construct a parser $\mathcal{P}_i$ for $A_i$. (1) For numeric attributes, $\mathcal{P}_i$ is a fixed mapping formula $i=\left\lceil \log_\gamma(x) \right\rceil$ that determines which pattern each parsed value $d$ corresponds to. (2) For string attributes, we use a prefix tree to store all patterns (i.e., regular expressions). Since different patterns can share several prefix tokens, their paths may overlap. This reduces the storage overhead of patterns and improves matching efficiency during the online phase.

\textbf{Patterns combination.} \textit{Mint} combines patterns of different attributes that appear together to form a span pattern, and assigns it a pattern ID (generates a UUID as pattern ID in our implementation). For example, if a span has two attributes, $A_1$ and $A_2$, and the pattern $P_{11}$ of $A_1$ always appears with the pattern $P_{23}$ of $A_2$, then $SP = [P_{11}, P_{23}]$ is a span pattern. \textit{Mint} store these span patterns in Pattern Library.

\subsubsection{\textit{Online Stage: Matching and Parsing.}}
When users employ \textit{Mint} for system tracing, \textit{Mint} performs online parsing on newly generated raw spans. 

\begin{figure}[t]
\centering
\includegraphics[width=\linewidth]{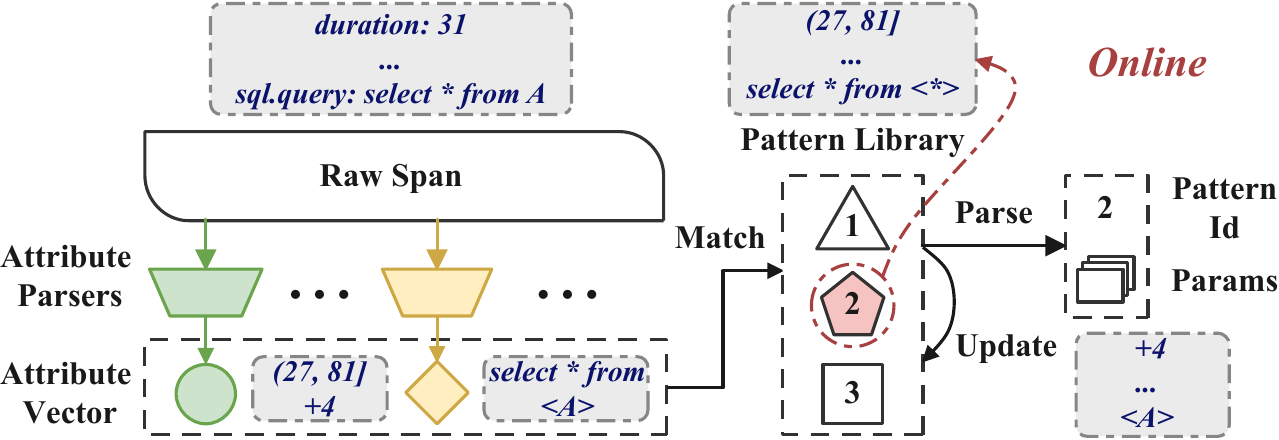}
\caption{The online stage of span parser.}
\label{fig:online}
\end{figure}

\textbf{Hierarchical attribute parsing.} As shown in Fig.~\ref{fig:online}, the core of the online parsing process is Hierarchical Attribute Parsing (HAP). \textit{Mint} parses each attribute of the incoming span in parallel. For each attribute: (1) it finds the matching pattern using the corresponding attribute parser (i.e., searching on the prefix tree or calculated through the mapping formula); (2) it uses the matched pattern as the common part and extracts the variable part based on the pattern. For string attributes, variables are extracted using regular expressions, while for numeric attributes, the difference from the interval's lower bound is calculated. If a new span pattern appears during the online stage (e.g., due to a system change), the corresponding parser updates to include the new pattern. We emphasize that the HAP process is highly parallel to meet the low-latency requirements at the online stage since different attribute parsers operate independently.

\textbf{Span pattern mapping.} After parsing each attributes of the incoming span, \textit{Mint} combines the parsed attribute patterns and matches the resulting span patterns with those in Pattern Library. If a consistent pattern is found, it returns the corresponding pattern ID. If not, it adds the new pattern to Pattern Library and assigns a new pattern ID, allowing Pattern Library to update based on the latest data.

Through the above steps, \textit{Mint} parses raw spans into patterns and parameters. Patterns are aggregated and stored in Pattern Library, while parameters are temporarily stored in Params Buffer until they are either transmitted or discarded.

\subsection{Inter-Trace Level Parsing}

\textbf{Sub-Trace construction.} After span parsing, \textit{Mint} agent links spans with the same trace ID by their parent IDs shown in Fig.~\ref{fig:trace_structure}, forming a sub-trace. 

Since a \textit{Mint} agent operates on an application node rather than the backend, its view is limited to the current node. Therefore, \textit{Mint} is designed to analyze the topology of trace segments (i.e., sub-traces) on the same node in real-time, rather than cross-node topology. This design prevents frequent waiting and interactions between agents, ensuring low latency for online tracing.

\begin{figure}[t]
\centering
\includegraphics[width=\linewidth]{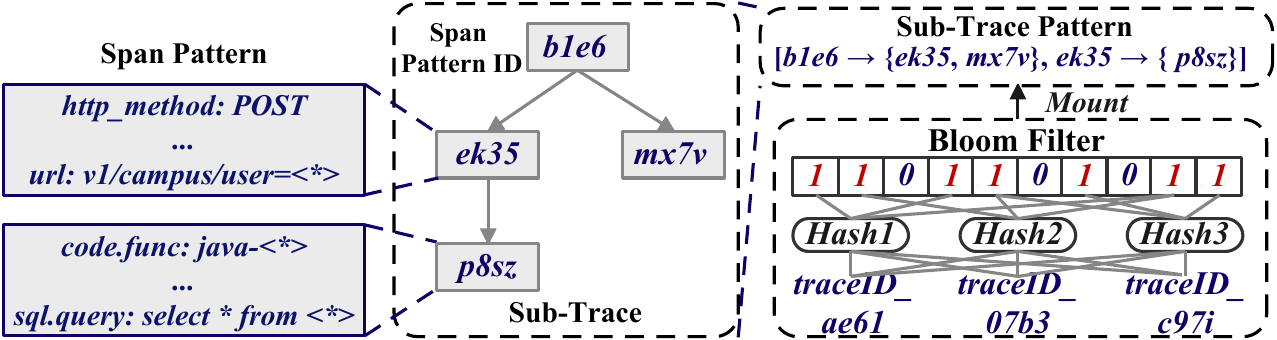}
\caption{\textit{Mint} uses a sub-trace pattern to store the topology information of a sub-trace. It also uses a Bloom Filter to efficiently store the trace metadata for each sub-trace pattern.}
\label{fig:topo}
\end{figure}

\textbf{Pattern extracting.} For an incoming sub-trace, \textit{Mint} encodes its topology information into a vector that captures the order and hierarchy of spans in the sub-trace as its pattern. Each dimension represents a parent-child relationship. For example, the encoded vector for the sub-trace in Fig.~\ref{fig:topo} is [b1e6 → \{ek35, mx7v\}, ek35 → \{p8sz\}]. It's important to note that each element in the sub-trace pattern is a span pattern ID, which corresponds to a span pattern. In other words, the sub-trace pattern includes both the topology information and the span contents information, as shown in Fig.~\ref{fig:topo}.

\textbf{Matching or updating.} For an incoming sub-trace's pattern, \textit{Mint} searches the Topo Pattern Library for an exact match. If a match is found, it is used as the matched pattern. If not, the pattern is added to the Topo Pattern Library as the matched pattern. By aggregating patterns, the topology information of traces with the same topology pattern only needs to be stored once.

\textbf{Metadata Mounting.} When querying, users typically request trace information using trace metadata (e.g., trace ID). Therefore, we need to record the matching relationship between trace metadata and its associated pattern. \textit{Mint} attaches a Bloom Filter to each sub-trace pattern, storing the metadata of all traces belonging to that pattern, as shown on the right side of Fig.~\ref{fig:topo}. A Bloom Filter is a highly space-efficient probabilistic data structure based on binary compression~\cite{bloom}, which is used to test whether an element is a member of a set. By using Bloom Filters, \textit{Mint} can determine which patterns a trace belongs to with low storage cost and high query efficiency. While Bloom Filters might falsely indicate that a trace belongs to a pattern, they will never miss a trace that does belong, ensuring trace coherence. Bloom Filter's false positives problem can be alleviated through upstream-downstream 
verification across multiple agents.


\subsection{Data Reporting}
We reiterate that through the above two levels of parsing, \textit{Mint} divides the trace data into three parts stored on the agent side: (1) span patterns and sub-trace patterns stored in Pattern Library; (2) trace metadata for retrieval stored in Bloom Filters and mounted on the corresponding sub-trace patterns; (3) variable parameters temporarily stored in Param Buffer. We now explain how \textit{Mint} agents filter and upload this information to backend for storage.

One of \textit{Mint}'s core design principles is to avoid completely discarding any trace, unlike the `1 or 0' paradigm. Therefore, \textit{Mint} agents periodically upload the full Pattern Library and Bloom Filters to the backend. This ensures that users can query and retrieve information for every trace without missing any one of them. We emphasize that due to widely existing commonality, millions of traces typically have only hundreds of patterns. As a result, the cost of storing patterns is significantly lower than storing raw trace data (approximately 0.5\% in our experiment).

For variable parameters, \textit{Mint} determines whether to send them to backend based on the importance of their associated traces. 
If a trace is marked as sampled, \textit{Mint} sends all variable parameters distributed across nodes to the backend. Thus \textit{Mint} backend can use parameters and patterns to reconstruct the complete information of sampled traces.

\textbf{Sampling rules.} Sampling rules determine which trace information should be fully retained. \textit{Mint} is compatible with existing sampling rules. Users can adopt head sampling by randomly marking some traces as sampled when requests are generated. Alternatively, they can mark all traces as sampled initially and filter them at the backend to apply tail sampling. 

Additionally, \textit{Mint} provides two samplers designed for `commonality + variability' paradigm, offering more comprehensive sampling. Similar to retroactive sampling~\cite{hindsight}, \textit{Mint}’s samplers perform biased sampling at the agent side. However, \textit{Mint} goes a step further by targeting not only symptomatic traces but also traces with rare execution paths. The design details of \textit{Mint}'s two samplers are shown in \S~\ref{subsec:mint_collector}.

\section{Implementation}
\begin{figure}[t]
\centering
\includegraphics[width=\linewidth]{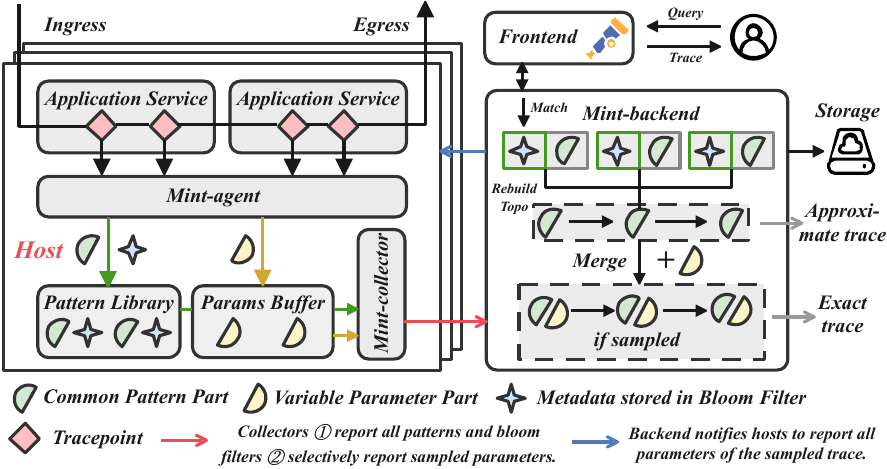}
\caption{Implementation of \textit{Mint} and use \textit{Mint} to capture and query for traces.}
\label{fig:imp_1}
\end{figure}

\textit{Mint} comprises several components, including multiple \textit{mint-agents} and \textit{mint-collectors} distributed on application hosts, a unified \textit{mint-backend} with a distributed trace storage engine, and a \textit{frontend} for query and visualization. Fig.~\ref{fig:imp_1} illustrates the implementation architecture of \textit{Mint}.

\subsection{Mint Agent}
We implemented \textit{mint-agent} in $\approx 2.5KLOC$ of Java. The \textit{mint-agent} supports various trace protocols (e.g., OpenTelemetry~\cite{opentelemetry}, Zipkin~\cite{zipkin}, and Jaeger~\cite{jaeger}) because \textit{Mint}'s subsequent parsing operations are decoupled from raw trace data generation. This allows it to support different protocols by changing the corresponding SDK (by default, we use the OpenTelemetry~\cite{opentelemetry} protocol in our implementation).

\textbf{Pattern Library.} Each \textit{mint-agent} allocates and maintains a Pattern Library in shared memory to store trace patterns and metadata stored in Bloom Filters. When the system is stable, the number of span and topology patterns converges, so the storage space for patterns does not increase. When the system changes, developers trigger \textit{Mint}’s reconstruct interface to rebuild the patterns since previous ones may become outdated. Bloom Filters grow with the number of traces, so \textit{Mint} pre-allocates a fixed-size buffer (default is 4 KB) for each Bloom filter. When the buffer is full, the Bloom filter is reported and reset. We implement the Bloom Filter using the Guava library~\cite{guava}, setting the `\textit{falsePositiveProbability}' parameter to $0.01$ by default.

\textbf{Params Buffer.} \textit{Mint-agent} reserves a fixed-size buffer (default 4 MB) in shared memory to temporarily store trace parameters. Params Buffer operates as a FIFO queue, with parameters from the same trace ID grouped into one block. Newly generated trace parameters blocks are added to the end of the queue. When the buffer is full, the block at the front of the queue is popped out.

\subsection{Mint Collector}
\label{subsec:mint_collector}
We implemented \textit{mint-collector} in $\approx 0.8KLOC$ of Java. The \textit{mint-collector} periodically (every 1 minute by default) reports patterns in Pattern Library and immediately reports Bloom Filters once they reach their size limit (on average, every 1.2 seconds). For parameters stored in the Params Buffer, \textit{Mint} uses two samplers designed for the `commonality + variability' paradigm to filter and decide whether to sample a trace (it can also accommodate other sampling rules like head sampling or tail sampling). Once a trace is marked as sampled, \textit{Mint} notifies collectors on all hosts through the backend to check and report all parameters of that trace, ensuring trace coherence.

\textbf{Symptom Sampler.} Symptom Sampler targets symptomatic traces by monitoring variable parameters in the Param Buffer and sampling the anomalies. For numerical parameters, it samples outliers exceeding the 95th percentile (P95). For string parameters, it samples values containing abnormal words, with the list of abnormal words being user-defined.

\textbf{Edge-Case Sampler.} Edge-Case Sampler targets traces with rare execution paths. It monitors the topology patterns in Pattern Library to track the number of traces matched to each topo pattern, increasing the sampling probability for less common traces. For example, if 99\% of traces follow pattern $A$ and 1\% follow pattern $B$, the Edge-Case Sampler will prioritize sampling traces that follow pattern $B$.

\subsection{Mint Backend and Querier}
\textit{Mint-backend} is implemented with a distributed trace storage engine and a querier. The distributed trace storage engine supports parallel and large-scale storage and querying of reported trace data. Notably, to make \textit{Mint}'s storage engine compatible with existing tracing frameworks, we design the data format of patterns (i.e., approximate traces) and parameters to be similar to traditional traces, and this storage method does not require decompression during queries.

\begin{figure}[t]
\centering
\includegraphics[width=\linewidth]{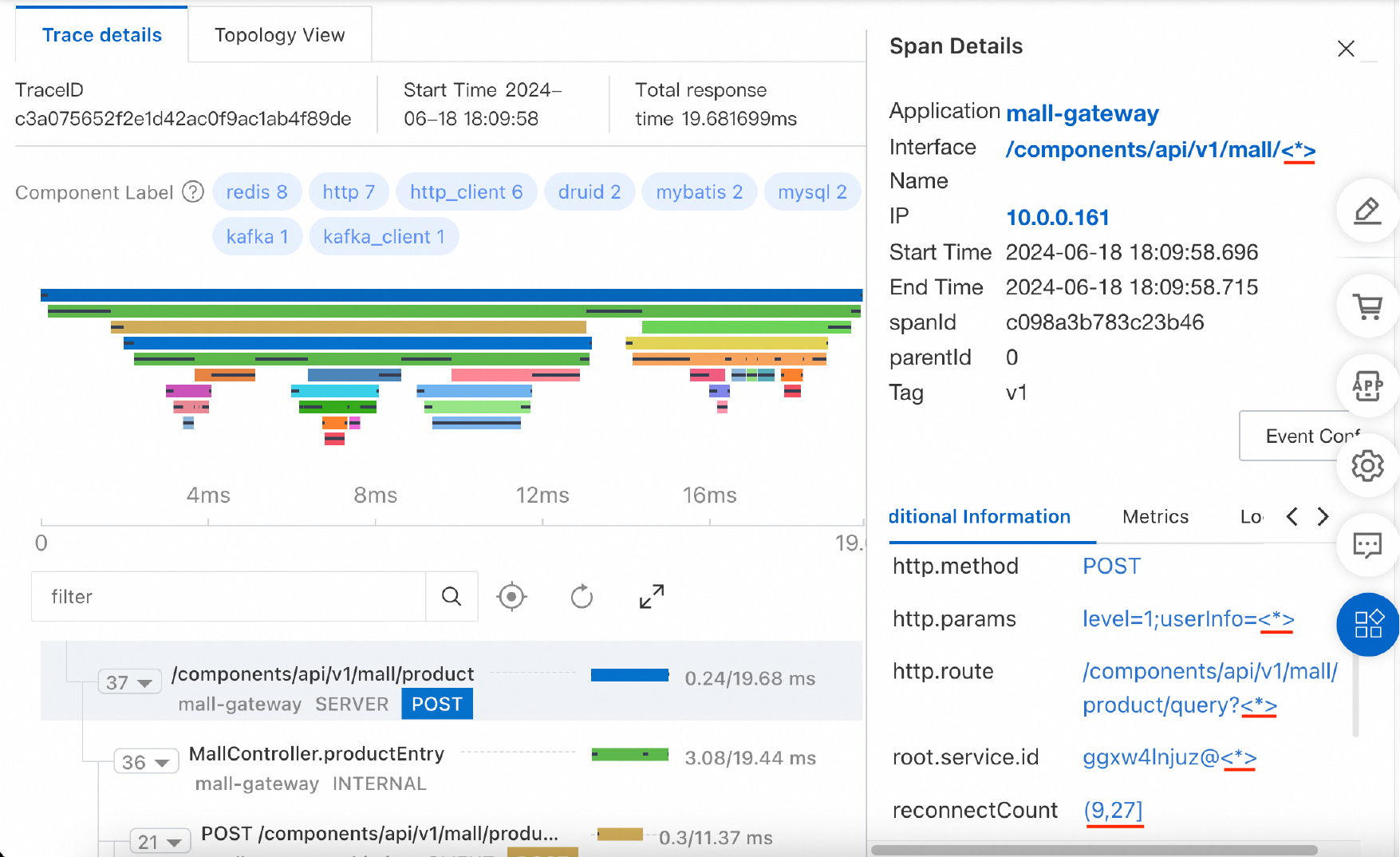}
\caption{An example of querying an unsampled trace to get an approximate trace, variables are masked by `<*>' and numbers are bucket-mapped.}
\label{fig:imp_2}
\end{figure}

\textbf{Query Logic.} When a user queries \textit{Mint} for trace information using trace metadata (e.g., trace ID), \textit{Mint} first checks each Bloom Filter for the presence of the trace's metadata. If found, it indicates that the pattern corresponding to the Bloom Filter is a segment (i.e., sub-trace) of the trace. \textit{Mint} then reconstructs these sub-traces into a complete approximate trace based on the matching relationships between the start and end operations of different segments. If the trace is marked as unsampled, the querier directly returns the approximate trace. If the trace is sampled, the exact parameters of the approximate trace have been sent to the backend.
Using these parameters and the approximate trace, the exact information of the queried trace can be reconstructed and returned. An example of querying an unsampled trace to get an approximate trace is shown as Fig~\ref{fig:imp_2}.

\section{Evaluation}
We now conduct experiments to evaluate \textit{Mint}, focusing on the following research questions.
\begin{itemize}
    \item How effective is \textit{Mint} in reducing trace data?
    \item How effective is \textit{Mint} in retaining more trace information (all traces capturing)?
    \item How much does \textit{Mint}'s commonality and variability analysis contribute to trace compression?
    \item What are the performance and scalability of \textit{Mint}?
\end{itemize}

\textbf{Benchmarks.} We used three distributed systems to evaluate \textit{Mint}. Two of them are open-source microservice benchmarks, which are widely used in previous trace analysis studies~\cite{Microrank, sieve}. The third is a real-world production microservice system from Alibaba, which allows us to better evaluate \textit{Mint} in a production environment.

OnlineBoutique (OB)~\cite{onlineboutique} is a web-based e-commerce application with 10 microservices implemented in various programming languages, communicating via gRPC. TrainTicket (TT)~\cite{trainticket} offers a railway ticketing service involving 45 services that communicate through synchronous REST invocations and asynchronous messaging. We deployed the OnlineBoutique and TrainTicket applications on a Kubernetes platform with 12 virtual machines. Each VM has an 8-core 2.10GHz CPU, 16GB of memory, and runs on Ubuntu 18.04.

\textbf{Baselines and implementation.} To better evaluate the effectiveness of \textit{Mint}, we compare it with four baseline tracing approaches, which are either widely used tracing frameworks or novel methods proposed in recent years.

\textcircled{\raisebox{-0.9pt}{1}} OpenTelemetry under head-sampling (\textbf{OT-Head})~\cite{ot-tailsampling}. 
In our implementation, we instrument all benchmark applications with the OpenTelemetry agent and collect trace data using the OpenTelemetry Collector, which is stored in Grafana Tempo and persisted to Elasticsearch~\cite{elasticsearch}. Unless otherwise specified, we set the head sampling rate to 5\%.

\textcircled{\raisebox{-0.9pt}{2}} OpenTelemetry under tail-sampling (\textbf{OT-Tail})~\cite{ot-headsampling}. 
OpenTelemetry's tail sampling strategy functions like a user-defined filter. To ensure its effectiveness, we tag all injected abnormal requests in the benchmark with an `\textit{is\_abnormal}' tag, allowing tail sampling to filter traces based on this tag.

\textcircled{\raisebox{-0.9pt}{3}} \textbf{Hindsight}~\cite{hindsight}. Hindsight is a tracing framework that implements retroactive sampling. Since Hindsight is compatible with OpenTelemetry, we configured the OpenTelemetry agent with Hindsight triggers on every application node. We used the same default parameters and configurations as specified in the Hindsight paper~\cite{hindsight}.

\textcircled{\raisebox{-0.9pt}{4}} \textbf{Sieve}~\cite{sieve}. Sieve is an online tail sampling approach that uses robust random cut forest (RRCF) to sample uncommon traces. We implemented it by using the OpenTelemetry agent and collector to generate traces and redirect them to the Sieve sampler for filtering and retention.

\subsection{Effectiveness in Reducing Trace Data}
In this experiment, we evaluate how effectively \textit{Mint} can reduce trace data. 
We measured the network and storage overhead of \textit{Mint} and four baseline tracing frameworks on OnlineBoutique and TrainTicket benchmarks. Additionally, we used OpenTelemetry with a 100\% sampling rate (\textbf{OT-Full}) as a reference for no trace reduction. To ensure fairness, we tag 5\% of the traffic injected into the benchmarks with an `\textit{is\_abnormal}' label and make all biased sampling methods to sample based on this field, this allows each tracing system to capture a consistent number of traces.

\begin{figure*}[t]
\centering
\includegraphics[width=\linewidth]{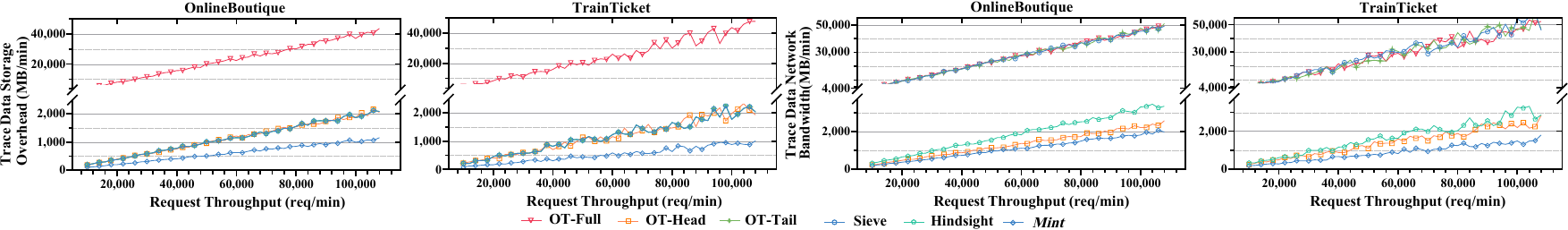}
\caption{Tracing network and storage overhead on OnlineBoutique and TrainTicket Benchmarks.}
\label{fig:exp_1}
\end{figure*} 

We measured the network overhead by monitoring the network bandwidth between application nodes and the tracing backend during the tracing process. To evaluate storage overhead, we measured the size of the trace data ultimately stored in Elasticsearch~\cite{elasticsearch} by the tracing backend. To assess the performance under different traffic loads, we conducted multiple experiments at varying request throughputs. 

Fig.~\ref{fig:exp_1} shows the results of our experiment. We can see that \textit{Mint} significantly reduces trace overhead in both network and storage compared to the baseline methods. Here is a further analysis of the results:

\textbf{OT-Head:} Head sampling randomly selects and retains sampled traces at the start of the trace's lifecycle. Therefore, its network and storage overhead is reduced to the sampling rate proportion (5\%) compared to OT-Full.

\textbf{OT-Tail \& Sieve:} Tail sampling decides and removes unsampled traces at the backend. As a result, it cannot reduce network overhead and remains similar to OT-Full, but it can reduce storage overhead to around the anomaly rate.

\textbf{Hindsight:} Hindsight performs biased sampling early at the agent side, which reduces both network and storage overhead. However, due to the need to transmit breadcrumbs, its network overhead is slightly higher than head sampling.

\textbf{\textit{Mint}}: \textit{Mint} reduces traces at the agent side, lowering both network and storage overhead. Additionally, \textit{Mint} optimizes trace storage by compressing traces based on commonality, further reducing trace data size. On average, \textit{Mint} reduces storage overhead to 2.7\% and network overhead to 4.2\%.

\subsection{Effectiveness in Retaining More Trace Information}
In this experiment, we demonstrate that \textit{Mint} can capture all requests and retain more trace information compared to current tracing frameworks with the same amount of trace data. We measure the quality of the trace information preserved by the tracing frameworks through two aspects: \textcircled{\raisebox{-0.9pt}{1}} Specific query response ability: We evaluate the ability of tracing frameworks to return the trace information in response to user queries, to determine if they can depict specific requests. \textcircled{\raisebox{-0.9pt}{2}} Analytical value of trace data: We assess the impact of the trace data produced by the tracing frameworks on downstream trace-based root cause analysis (RCA) methods, evaluating analytical value of captured trace data.

When evaluating the effectiveness of retained trace information, we ensure fairness by controlling the size of the trace data saved by tracing frameworks to be consistent. We set the budget trace reduction rate to 5\%, meaning each framework's final saved trace data is 5\% of the original size. 

\textbf{Query Response Ability.} To reflect real-world user query behavior for trace data, we randomly selected three subsystems from Alibaba for continuous monitoring over a period of time (14 days). During this monitoring period, we used the OpenTelemetry collector to collect all request data and redirected it to the tracing frameworks under evaluation (\textit{Mint} and four baselines) to complete subsequent trace data processing steps. Simultaneously, we recorded which trace IDs users queried daily during this period and applied these queries to the tracing frameworks. If the tracing framework could return the complete information of the queried trace, it was marked as an `\textit{exact hit}'. If it could return approximate information of the queried trace, it was marked as a `\textit{partial hit}'. If there was no record at all, it was marked as a `\textit{miss}'.

\begin{figure}[t]
\centering
\includegraphics[width=\linewidth]{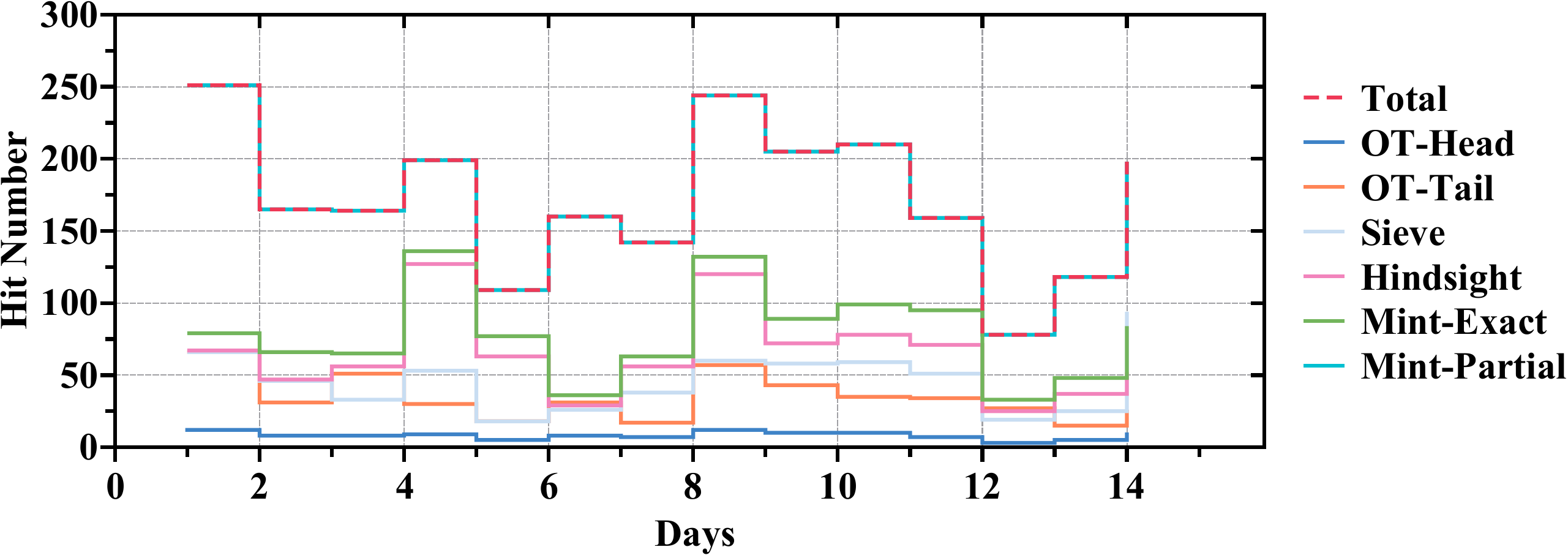}
\caption{Hit number for user queries in Alibaba during 14 days, demonstrating \textit{Mint} can respond to all requests.}
\label{fig:exp_2}
\end{figure}

Fig.~\ref{fig:exp_2} shows the number of hits each tracing framework achieved during the experiment period in response to queries. The red dashed line, labeled `Total', represents the total number of user queries per day during this period. As seen, when considering partial hits, \textit{Mint} responds to all queries, meaning it can provide at least approximate information for every trace. When considering only exact hits, \textit{Mint} still outperforms baseline methods by responding to more queries.

\input{table/rq2_desp}

\input{table/RCA}

\textbf{Effectiveness for downstream analysis.} To simulate real-world microservices problem analysis, we conducted chaos engineering on the OnlineBoutique and TrainTicket benchmarks. We used Chaosblade~\cite{Chaosblade} to inject a total of 56 faults into these two benchmark microservices. 
The types of injected faults are shown in Table~\ref{tab:rq2_desp}.
Notably, the two open-source microservice benchmarks we used are widely adopted in previous trace analysis studies~\cite{Microrank, sieve, trastrainer}, and the injected failures are also among the most common types found in real-world production environments~\cite{Microrank}.

We deployed \textit{Mint} and the other four baselines on the benchmark microservices to capture trace data. Using three classic trace-based analysis methods (i.e., MicroRank~\cite{Microrank}, TraceRCA~\cite{TraceRCA}, TraceAnomaly~\cite{TraceAnomaly}), we conducted RCA based on the captured trace data. We then calculated the top-1 accuracy (A@1)~\cite{TraceRCA} of the analysis results.

Table~\ref{tab:rca_res} shows the A@1 for different combinations of tracing frameworks and RCA methods. It can be seen that \textit{Mint} significantly improves the accuracy of downstream root cause analysis compared to baseline methods. 

MicroRank~\cite{Microrank} and TraceRCA~\cite{TraceRCA} require a sufficient number of common-case traces to conduct spectrum analysis~\cite{spectrum} for root cause identification. TraceAnomaly~\cite{TraceAnomaly} compares the abnormal trace with normal templates to locate root causes, also needing enough common traces to establish normal templates. The previous `1 or 0' sampling strategy, which entirely discarded common traces, severely weakened these RCA methods, resulting in A@1 below 38\%. 

Although using the same trace storage size, \textit{Mint} with the `commonality + variability' approach retains essential information for all traces and detailed information of edge cases, fully enhancing the performance of RCA methods, with an average increase of A@1 from 25\% to 50\%.

\subsection{Contribution of Commonality and Variability Analysis}
\label{subsubsec:expr_rq3}

In \textit{RQ1}, we have explored \textit{Mint}'s overall trace reduction capability,
in this experiment, we now focus on \textit{Mint}'s lossless compression ability.
We compare \textit{Mint}'s compression ratio with other compression tools. Notably, trace compression in this context requires that compressed data can be directly used for retrieval and query without decompression. General-purpose compressors (e.g., gzip~\cite{gzip},
bzip2~\cite{bzip2}) are unsuitable here because they compress the original file character-by-character into a binary file, making direct retrieval or query impossible without decompressing the entire dataset first. This does not practically reduce storage costs and significantly increases query latency. Therefore, we compare \textit{Mint} with log-specific compressors (i.e., logzip~\cite{liu2019logzip}, logreducer~\cite{logreducer21fast} and CLP~\cite{CLP}), which also eliminate redundancy based on data characteristics for compression.

\begin{figure}[t]
\centering
\includegraphics[width=\linewidth]{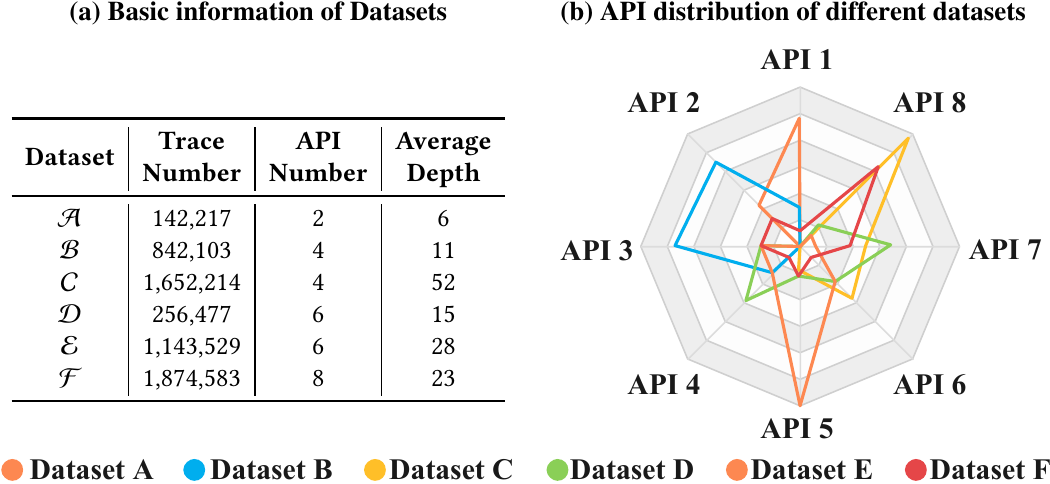}
\caption{Description of 6 datasets in Alibaba.}
\label{fig:exp_3_desp}
\end{figure} 

Additionally, to individually evaluate \textit{Mint}'s effectiveness at both levels of commonality and variability parsing, we perform an ablation study and design two variants: \textit{Mint} without inter-span level parsing (denoted as w/o $\mathcal{S}p$) and \textit{Mint} without inter-trace level parsing (denoted as w/o $\mathcal{T}p$).

To evaluate \textit{Mint}'s trace compression ability in various production environments, we selected 6 subsystems from Alibaba to generate real-world traces, each with different API counts and call depths. Detailed descriptions of these datasets are shown in Fig.~\ref{fig:exp_3_desp}.

Table~\ref{tab:expr_rq3} shows the compression ratios of the five approaches tested on the six datasets. \textit{Mint} outperformed the two baseline methods by an average of 14.90 to 28.38 in compression ratio. This improvement is because \textit{Mint} more effectively considers the unique characteristics of trace data and compresses traces based on topology, achieving higher performance. Additionally, \textit{Mint} significantly outperforms its two ablation variants, with an average improvement of 8.45 to 26.45 in compression ratio, demonstrating that both inter-span and inter-trace level parsing contribute to trace compression.


\input{table/rq3}

\subsection{Mint Overhead and Scalability}
\label{subsec:overhead}
\textbf{End-to-End Tracing Overhead.} To ensure \textit{Mint} is a practical tool, we evaluate and demonstrate that \textit{Mint} produces acceptable computational, network, and storage overhead. We also assess the impact of request throughput and request complexity to test \textit{Mint}'s scalability. To ensure fairness in experiments, both trace data generated by \textit{Mint} and by all baselines were persisted to Elasticsearch~\cite{elasticsearch}.
We conduct our experiments using a production microservice system from Alibaba, which includes web services, MongoDB~\cite{mongodb} access, and MySQL~\cite{mysql} access. We create three identical replicas, each with \textit{Mint}, OpenTelemetry (with head sampling), or no tracing framework installed. To ensure fairness, we control \textit{Mint} and OpenTelemetry to have the same sampling rate (10\%). During the experiments, we inject traffic into the three replicas and conduct 14 load tests, with varying request throughput and API requests. Engineers indicate that load tests of this scale are able to reflect Mint's performance during peak traffic periods in most systems.

\begin{figure}[t]
\centering
\includegraphics[width=\linewidth]{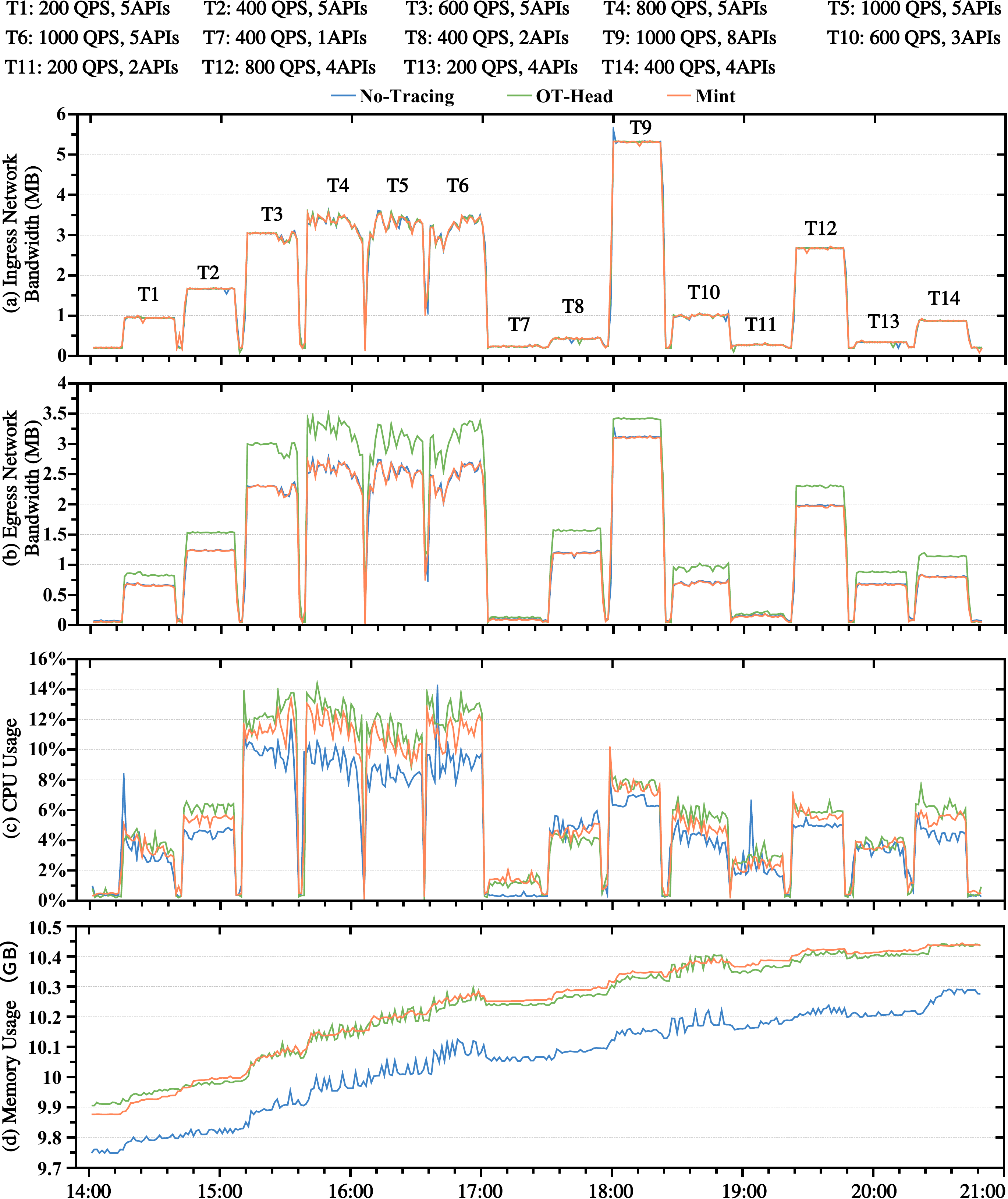}
\caption{Tracing overhead during 14 load tests on Alibaba's production microservices system.}
\label{fig:exp_4}
\end{figure} 

The experimental results are shown in Fig.~\ref{fig:exp_4}. Fig.~\ref{fig:exp_4} (a) indicates that all three replicas received the same traffic during the 14 tests. As seen in Fig.~\ref{fig:exp_4} (b), \textit{Mint} effectively reduced the trace data traffic through compression, with egress network bandwidth increasing by only 2.88\% compared to no tracing. In contrast, OT-Head increased the bandwidth by 19.35\%. Fig.~\ref{fig:exp_4} (c) shows that \textit{Mint}'s computational overhead during tracing is acceptable, with an average CPU usage increase of 0.86\% compared to no tracing, and 0.39\% less than OT-Head. Fig.~\ref{fig:exp_4} (d) illustrates that \textit{Mint}'s storage overhead is also acceptable, similar to OT-Head, with an average increase of 1.8\% compared to no tracing.

\begin{figure}[t]
\centering
\includegraphics[width=\linewidth]{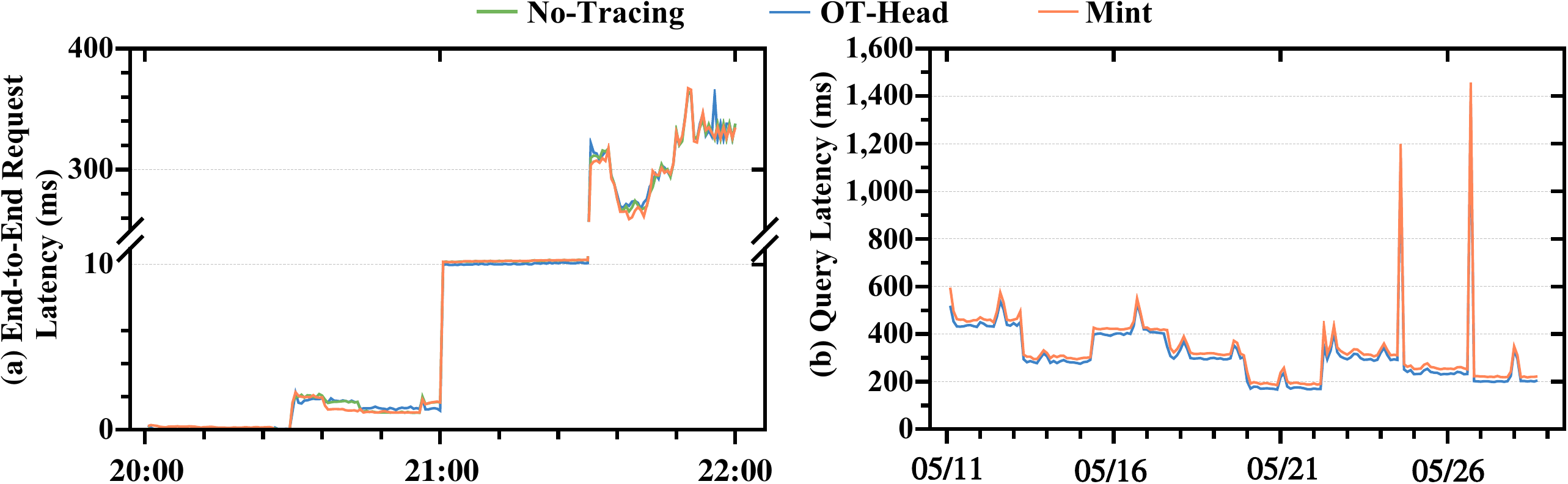}
\caption{End-to-End request latency and query latency on Alibaba's production microservices system.}
\label{fig:exp_42}
\end{figure} 

\textbf{Latency.} We also evaluated the impact of using \textit{Mint} for tracing on end-to-end request latency and the latency of querying traces with \textit{Mint}. The experiment was conducted on the same production system of Alibaba as in the tests in Fig.~\ref{fig:exp_4}. Fig.~\ref{fig:exp_42} (a) shows that for different types of requests, using \textit{Mint} increased the request latency by an average of 0.21\%, which is entirely acceptable. Fig.~\ref{fig:exp_42} (b) indicates that using \textit{Mint} for querying took, on average, 4.2\% longer than using OpenTelemetry, with the P95 latency below 1 second, meeting production environment requirements.

\input{table/parse_1}

\textbf{Pattern extraction performance.} In the previous experiments, we verified \textit{Mint}'s end-to-end performance. Next, we designed an experiment to test the pattern extraction capabilities of the Span Parser and Trace Parser. We collected raw trace data generated by five sub-services in Alibaba Cloud over an hour. We used Span Parser and Trace Parser to extract patterns at the span level and trace level, respectively, and recorded the number of patterns obtained, as shown in Table~\ref{tab:parser}. The results show that both Span Parser and Trace Parser effectively aggregate patterns from trace data of different sub-services. The compression ratio from the original log count to span-level patterns ranges from 6,681 to 13,362, and from the original log count to trace level patterns, it ranges from 15,780 to 30,842.

\section{Discussions}
\subsection{Parameter Sensitivity}
\begin{figure}[t]
\centering
\includegraphics[width=0.85\linewidth]{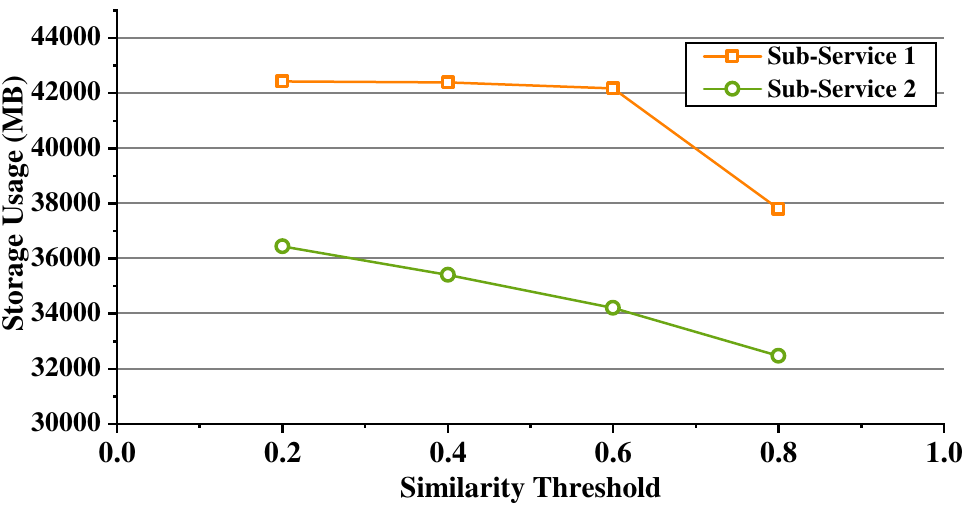}
\caption{The total storage size of patterns and parameters with the similarity threshold at 0.2, 0.4, 0.6, and 0.8.}
\label{fig:param}
\end{figure}

The default parameter settings are based on our real-world implementation, where we found them to deliver the best overall results. The primary parameter in \textit{Mint} is the similarity threshold in the Span Parser. A higher threshold leads to more patterns but fewer parameters. We used raw trace data from two sub-services, as mentioned in \S~\ref{subsec:overhead}, and set the similarity threshold at 0.2, 0.4, 0.6, and 0.8 to explore the effect on the total storage size of patterns and parameters (without sampling or compression). As shown in Fig.~\ref{fig:param}, the total storage size for patterns and parameters decreases as the similarity threshold increases. However, an excessively high similarity threshold reduces the differences between spans within the same pattern, weakening the parameter extraction effectiveness. Considering both total storage size and parameter extraction effectiveness, we set the default similarity threshold to 0.8.
In most cases, using the default settings will yield satisfactory results.

\subsection{Trace Coherence}
Trace coherence refers to the requirement that a distributed trace should maintain its topological integrity, as a fragmented trace loses its end-to-end visibility necessary for analysis~\cite{hindsight}. \textit{Mint} ensures that all stored trace (whether it’s an unsampled trace or a sampled trace) maintains trace coherence. 
\textit{Mint} guarantees trace coherence through several designs: (1) All trace topology patterns are recorded at every node they pass through and fully reported to the backend. (2) The no-miss property of Bloom Filters ensures that \textit{Mint} can gather all segments of a queried trace during a query. (3) Each trace topology segment's entry and exit operations identify its upstream and downstream segments, respectively. By upstream-downstream matching, the complete topology of each trace can be reconstructed.

\subsection{Production Use Cases of Approximate Traces}
A core feature of \textit{Mint} is that it does not discard unsampled traces. Instead, it returns an approximate trace. We analyzed two use cases (UCs) where approximate traces were used for analysis after \textit{Mint} was deployed in a real-world system in Alibaba Cloud. This demonstrates that approximate traces of regular traces are highly beneficial for trace analysis.

\textbf{Trace Exploration (UC 1).} 
Take the real-world case from \S~\ref{subsubsec:RQ2} as an example. On Mar. 25, 2024, when users needed to explore and visualize traces that were unsampled but essential for analysis, previous tracing frameworks based on the `1 or 0' strategy provided no results due to discarding the unsampled traces. However, with \textit{Mint}, approximately 80\% of the Trace Explorer functionality can be remained using approximate traces. This includes the full trace execution path, flame graph, types and approximate content of each operation. Using above information, users analyzed the topology and key data of these traces, meeting their analysis needs for the queries in this case.

\textbf{Batch Trace Analysis (UC 2).} 
In production scenarios, analysts often need to perform batch analysis on traces over a specific period or with certain characteristics. For example, they may need to plot scatter diagrams of trace features or aggregate the topology of multiple traces. Before using \textit{Mint}, sampling limitations meant that an average batch query could only retrieve 3,000–5,000 spans, which was insufficient for thorough analysis, as this might only cover a few minutes of data. With \textit{Mint}, unsampled traces can be included in batch analysis through approximate traces, which still retain execution paths and basic attributes, increasing the number of spans available to millions. This expanded data volume allows users to achieve better batch processing results and improved accuracy.

\section{Conclusions}
In this paper, we propose the `commonality + variability' paradigm for trace reduction and design \textit{Mint} to implement this paradigm, capturing full requests in a cost-efficient way. \textit{Mint} parses trace data on two levels, aggregates patterns, and filters parameters, ensuring effective recording of each request. Our experiments demonstrate that \textit{Mint} retains more useful information while reducing trace volume and is lightweight enough to meet production requirements. We also implemented \textit{Mint} at a large cloud provider, Alibaba, where user feedback indicated a significant improvement in user experience and facilitated further analysis.

\begin{acks}
We sincerely appreciate the insightful feedback from the anonymous reviewers and our shepherd, Ryan Huang. This work was supported  by the Guangdong Basic and Applied Basic Research Foundation under Grant 2023B1515020054, in part by the National Natural Science Foundation of China under Grant 62272495, and in part by Alibaba project under Grant 20925056. The corresponding author is Pengfei Chen.
\end{acks}

\bibliographystyle{plain}
\bibliography{references.bib}

\end{document}

%% file: table/empi_rq3.tex
\begin{table}[t]
\caption{The statistics of occurrence and proportion of  commonality of traces from three services in company $\mathcal{A}$ (\textbf{\#} denotes occurrence and \textbf{\%} denotes proportion).}
\centering
\resizebox{\linewidth}{!}{
\begin{threeparttable}
\begin{tabular}{ c | c  c | c c | c  c }
\toprule                                                               
\multirow{2}{*}{Level} & \multicolumn{2}{c|}{Service A} & \multicolumn{2}{c|}{Service B} &  \multicolumn{2}{c}{Service C} \\ \cline{2-7}
                         & \textbf{\#}     & \textbf{\%}
                         & \textbf{\#}     & \textbf{\%} 
                         & \textbf{\#}     & \textbf{\%}  \\
\midrule

Inter-trace &  145,701 & \textbf{48.20} & 169,701 & \textbf{56.14} & 104,101 & \textbf{34.44}    \\
Inter-span  &   3,382,701 & \textbf{35.31} & 5,302,701 & \textbf{45.34} & 2,448,701 & \textbf{25.55}    \\
\bottomrule
\end{tabular} 
\end{threeparttable}}
\label{tab:empi_rq3}
\end{table}

%% file: table/rq2_desp.tex
\begin{table}[t]
\caption{Choice of injected faults and downstream methods}
\centering
\resizebox{0.8\linewidth}{!}{
\begin{threeparttable}
\begin{tabular}{c}
\toprule
\rowcolor[HTML]{dae8fc} 
\textbf{Injected Fault Types}                                                                                                                                           \\
CPU exhaustion, memory exhaustion, \\
network delays, code exceptions, error returns                                                                                                     \\
\rowcolor[HTML]{dae8fc} 
\textbf{Trace-based RCA Methods}       \\
MicroRank~\cite{Microrank}, TraceRCA~\cite{TraceRCA}, TraceAnomaly~\cite{TraceAnomaly} \\
 \bottomrule
\end{tabular}
\end{threeparttable}}
\label{tab:rq2_desp}
\end{table}

%% file: table/RCA.tex
\begin{table}[t]
\caption{Comparison of the effects of different tracing frameworks in downstream root cause analysis's accuracy.}
\centering
\resizebox{0.95\linewidth}{!}{
\begin{threeparttable}
\begin{tabular}{c | c | c  c  c  c c}
\toprule       
\multirow{2}{*}{\begin{tabular}[c]{@{}c@{}} \textbf{Bench-}\\ \textbf{mark}\end{tabular}} & \multirow{2}{*}{\begin{tabular}[c]{@{}c@{}} \textbf{RCA}\\ \textbf{Method}\end{tabular}} & \multicolumn{5}{c}{\textbf{Tracing Framework}} \\ \cmidrule{3-7}
& & \textbf{OT-Head} & \textbf{OT-Tail} & \textbf{Sieve} & \textbf{Hindsight} & \textbf{\textit{Mint}} \\
\midrule

\multirow{3}{*}{OB}& MicroRank & 0.1563 & 0.2188 & 0.2813 & 0.2188 & \textbf{0.6563}  \\
& TraceAnomaly & 0.2813 & 0.2500 & 0.3750 & 0.3438 & \textbf{0.7037}  \\
& TraceRCA & 0.2500 & 0.2500 & 0.3438 & 0.2188 & \textbf{0.6563}  \\
\midrule

\multirow{3}{*}{TT}& MicroRank & 0.0714 & 0.1429 & 0.1786 & 0.1786 & \textbf{0.5357}  \\
& TraceAnomaly & 0.1786 & 0.1786 & 0.2857 & 0.3214 & \textbf{0.5714}  \\
& TraceRCA & 0.1429 & 0.1786 & 0.2500 & 0.1429 & \textbf{0.5000}  \\

\bottomrule
\end{tabular} 
\end{threeparttable}}
\label{tab:rca_res}
\end{table}

%% file: table/rq3.tex
\begin{table}[t]
\caption{Comparison in terms of Compression Ratio.}
\centering
\resizebox{0.9\linewidth}{!}{
\begin{threeparttable}
\begin{tabular}{ c c  c c c c c}
\toprule                                                               
\textbf{Dataset} & \textbf{LogZip} & \textbf{LogReducer} & \textbf{CLP}  & \textbf{w/o $\mathcal{S}p$} & \textbf{w/o $\mathcal{T}p$} & \textbf{\textit{Mint}}
    \\
\midrule

     $\mathcal{A}$     & 16.7989    & 19.9594  & 22.7130    & 21.2503    & 23.1391    & \textbf{45.1874}   \\

 $\mathcal{B}$      & 13.0634    & 10.2291   &  14.0553    & 14.3892    & 15.9906    & \textbf{41.0603}             \\

 $\mathcal{C}$      & 5.2411    & 7.8613   & 11.5995    & 14.3229    & 13.7895    & \textbf{22.7690}          \\

 $\mathcal{D}$      & 11.0920    & 11.4943  & 14.4578    & 10.2255    & 18.1101    & \textbf{36.6724}       \\

 $\mathcal{E}$      & 8.7774    & 9.0126  & 12.1723    & 10.1943    & 17.1917    & \textbf{32.0245}        \\

 $\mathcal{F}$      & 9.2336    & 10.6611  & 15.3990    & 8.9231    & 19.7713    & \textbf{29.7024} \\



\bottomrule
\end{tabular} 
\end{threeparttable}}
\label{tab:expr_rq3}
\end{table}

%% file: table/parse_1.tex
\begin{table}[t]
\caption{Pattern extraction results of Span Parser and Trace Parser on 5 sub-services in Alibaba Cloud.}
\centering
\resizebox{\linewidth}{!}{
\begin{threeparttable}
\begin{tabular}{ c | c | c | c }
\toprule                                                               
\textbf{Sub-Service}                    
& \begin{tabular}[c]{@{}c@{}} \textbf{Raw Trace}\\ \textbf{Number}\end{tabular}
& \begin{tabular}[c]{@{}c@{}} \textbf{Span Level}\\ \textbf{Pattern Number}\end{tabular}    & \begin{tabular}[c]{@{}c@{}} \textbf{Trace Level }\\ \textbf{Pattern Number}\end{tabular}     \\    
\midrule

     $\mathcal{S}1$     & 146,985    & 11      & 8              \\

 $\mathcal{S}2$     & 126,245    & 10      & 8              \\

 $\mathcal{S}3$     & 93,546    & 14      & 5              \\

 $\mathcal{S}4$     & 92,527    & 7      & 3              \\

 $\mathcal{S}5$     & 79,179    & 9      & 3              \\



\bottomrule
\end{tabular} 
\end{threeparttable}}
\label{tab:parser}
\end{table}

%% file: main.bbl
\begin{thebibliography}{10}

\bibitem{zipkin}
Zipkin homepage.
\newblock \url{https://zipkin.io}, 2023.
\newblock [Online].

\bibitem{lzma}
7za tool.
\newblock \url{https://linux.die.net/man/1/7za}, 2024.
\newblock [Online].

\bibitem{bzip2}
The bzip2 homepage.
\newblock \url{https://sourceware.org/bzip2/}, 2024.
\newblock [Online].

\bibitem{gzip}
The gzip homepage.
\newblock \url{ https://www.gzip.org}, 2024.
\newblock [Online].

\bibitem{alibaba}
Alibaba.
\newblock Alibaba.
\newblock \url{https://www.alibaba.com/}, 2024.
\newblock Accessed Oct. 7, 2024.

\bibitem{magpie}
Paul Barham, Rebecca Isaacs, Richard Mortier, and Dushyanth Narayanan.
\newblock Magpie: Online modelling and performance-aware systems.
\newblock In {\em 9th Workshop on Hot Topics in Operating Systems (HotOS IX)},
  Lihue, HI, May 2003. USENIX Association.

\bibitem{Chaosblade}
Chaosblade.
\newblock Chaosblade.
\newblock \url{https://github.com/chaosblade-io/chaosblade}, 2024.
\newblock Accessed Mar. 6, 2024.

\bibitem{logarchive}
Robert Christensen and Feifei Li.
\newblock Adaptive log compression for massive log data.
\newblock In {\em Proceedings of the 2013 ACM SIGMOD International Conference
  on Management of Data}, SIGMOD '13, page 1283–1284, New York, NY, USA,
  2013. Association for Computing Machinery.

\bibitem{datadog}
datadog.
\newblock What is distributed tracing? how it works \& use cases.
\newblock
  \url{https://www.datadoghq.com/knowledge-center/distributed-tracing/}, 2024.
\newblock Accessed: 2024/4/29.

\bibitem{elise}
Hailun Ding, Shenao Yan, Juan Zhai, and Shiqing Ma.
\newblock {ELISE}: A storage efficient logging system powered by redundancy
  reduction and representation learning.
\newblock In {\em 30th USENIX Security Symposium (USENIX Security 21)}, pages
  3023--3040. USENIX Association, August 2021.

\bibitem{elasticsearch}
Elasticsearch.
\newblock Elasticsearch.
\newblock \url{https://github.com/elastic/elasticsearch}, 2024.
\newblock Accessed Mar. 6, 2024.

\bibitem{mlc}
Bo~Feng, Chentao Wu, and Jie Li.
\newblock Mlc: An efficient multi-level log compression method for cloud backup
  systems.
\newblock In {\em 2016 IEEE Trustcom/BigDataSE/ISPA}, pages 1358--1365, 2016.

\bibitem{x-trace}
Rodrigo Fonseca, George Porter, Randy~H. Katz, and Scott Shenker.
\newblock {X-Trace}: A pervasive network tracing framework.
\newblock In {\em 4th USENIX Symposium on Networked Systems Design \&
  Implementation (NSDI 07)}, Cambridge, MA, April 2007. USENIX Association.

\bibitem{trainticket}
FudanSELab.
\newblock Trainticket.
\newblock \url{https://github.com/FudanSELab/train-ticket}, 2024.
\newblock Accessed June. 6, 2024.

\bibitem{Sage}
Yu~Gan, Mingyu Liang, Sundar Dev, David Lo, and Christina Delimitrou.
\newblock Sage: Practical and scalable ml-driven performance debugging in
  microservices.
\newblock In {\em Proceedings of the 26th ACM International Conference on
  Architectural Support for Programming Languages and Operating Systems},
  ASPLOS '21, page 135–151, New York, NY, USA, 2021. Association for
  Computing Machinery.

\bibitem{gan2023sleuth}
Yu~Gan, Guiyang Liu, Xin Zhang, Qi~Zhou, Jiesheng Wu, and Jiangwei Jiang.
\newblock Sleuth: A trace-based root cause analysis system for large-scale
  microservices with graph neural networks.
\newblock In {\em Proceedings of the 28th ACM International Conference on
  Architectural Support for Programming Languages and Operating Systems, Volume
  4}, pages 324--337, 2023.

\bibitem{samplehst}
Alim~Ul Gias, Yicheng Gao, Matthew Sheldon, José~A. Perusquía, Owen O'Brien,
  and Giuliano Casale.
\newblock Samplehst: Efficient on-the-fly selection of distributed traces,
  2022.

\bibitem{onlineboutique}
GoogleCloudPlatform.
\newblock Onlineboutique.
\newblock \url{https://github.com/GoogleCloudPlatform/microservices-demo},
  2024.
\newblock Accessed June. 6, 2024.

\bibitem{guava}
guava.
\newblock guava.
\newblock \url{https://github.com/google/guava}, 2024.
\newblock Accessed Mar. 6, 2024.

\bibitem{GMTA}
Xiaofeng Guo, Xin Peng, Hanzhang Wang, Wanxue Li, Huai Jiang, Dan Ding, Tao
  Xie, and Liangfei Su.
\newblock Graph-based trace analysis for microservice architecture
  understanding and problem diagnosis.
\newblock In {\em Proceedings of the 28th ACM Joint Meeting on European
  Software Engineering Conference and Symposium on the Foundations of Software
  Engineering}, ESEC/FSE 2020, page 1387–1397, New York, NY, USA, 2020.
  Association for Computing Machinery.

\bibitem{trastrainer}
Haiyu Huang, Xiaoyu Zhang, Pengfei Chen, Zilong He, Zhiming Chen, Guangba Yu,
  Hongyang Chen, and Chen Sun.
\newblock Trastrainer: Adaptive sampling for distributed traces with system
  runtime state.
\newblock {\em Proc. ACM Softw. Eng.}, 1(FSE), July 2024.

\bibitem{tprof}
Lexiang Huang and Timothy Zhu.
\newblock Tprof: Performance profiling via structural aggregation and automated
  analysis of distributed systems traces.
\newblock In {\em Proceedings of the ACM Symposium on Cloud Computing}, SoCC
  '21, page 76–91, New York, NY, USA, 2021. Association for Computing
  Machinery.

\bibitem{sieve}
Zicheng Huang, Pengfei Chen, Guangba Yu, Hongyang Chen, and Zibin Zheng.
\newblock Sieve: Attention-based sampling of end-to-end trace data in
  distributed microservice systems.
\newblock In {\em 2021 IEEE International Conference on Web Services (ICWS)},
  pages 436--446, 2021.

\bibitem{jaeger}
jaeger.
\newblock Jaeger.
\newblock \url{https://www.jaegertracing.io/}, 2023.
\newblock Accessed: 2024/4/29.

\bibitem{kaldor2017canopy}
Jonathan Kaldor, Jonathan Mace, Micha{\l} Bejda, Edison Gao, Wiktor Kuropatwa,
  Joe O'Neill, Kian~Win Ong, Bill Schaller, Pingjia Shan, Brendan Viscomi,
  et~al.
\newblock Canopy: An end-to-end performance tracing and analysis system.
\newblock In {\em Proceedings of the 26th symposium on operating systems
  principles}, pages 34--50, 2017.

\bibitem{facebook}
Jonathan Kaldor, Jonathan Mace, Micha\l{} Bejda, Edison Gao, Wiktor Kuropatwa,
  Joe O'Neill, Kian~Win Ong, Bill Schaller, Pingjia Shan, Brendan Viscomi,
  Vinod Venkataraman, Kaushik Veeraraghavan, and Yee~Jiun Song.
\newblock Canopy: An end-to-end performance tracing and analysis system.
\newblock In {\em Proceedings of the 26th Symposium on Operating Systems
  Principles}, SOSP '17, page 34–50, New York, NY, USA, 2017. Association for
  Computing Machinery.

\bibitem{milliScope}
Chien-An Lai, Josh Kimball, Tao Zhu, Qingyang Wang, and Calton Pu.
\newblock milliscope: A fine-grained monitoring framework for performance
  debugging of n-tier web services.
\newblock In {\em 2017 IEEE 37th International Conference on Distributed
  Computing Systems (ICDCS)}, pages 92--102, 2017.

\bibitem{wset}
Pedro Las-Casas, Jonathan Mace, Dorgival Guedes, and Rodrigo Fonseca.
\newblock Weighted sampling of execution traces: Capturing more needles and
  less hay.
\newblock In {\em Proceedings of the ACM Symposium on Cloud Computing}, SoCC
  '18, page 326–332, New York, NY, USA, 2018. Association for Computing
  Machinery.

\bibitem{sifter}
Pedro Las-Casas, Giorgi Papakerashvili, Vaastav Anand, and Jonathan Mace.
\newblock Sifter: Scalable sampling for distributed traces, without feature
  engineering.
\newblock In {\em Proceedings of the ACM Symposium on Cloud Computing}, SoCC
  '19, page 312–324, New York, NY, USA, 2019. Association for Computing
  Machinery.

\bibitem{TraceRCA}
Zeyan Li, Junjie Chen, Rui Jiao, Nengwen Zhao, Zhijun Wang, Shuwei Zhang,
  Yanjun Wu, Long Jiang, Leiqin Yan, Zikai Wang, Zhekang Chen, Wenchi Zhang,
  Xiaohui Nie, Kaixin Sui, and Dan Pei.
\newblock Practical root cause localization for microservice systems via trace
  analysis.
\newblock In {\em 2021 IEEE/ACM 29th International Symposium on Quality of
  Service (IWQOS)}, pages 1--10, 2021.

\bibitem{cowic}
Hao Lin, Jingyu Zhou, Bin Yao, Minyi Guo, and Jie Li.
\newblock Cowic: A column-wise independent compression for log stream analysis.
\newblock In {\em 2015 15th IEEE/ACM International Symposium on Cluster, Cloud
  and Grid Computing}, pages 21--30, 2015.

\bibitem{microhecl}
Dewei Liu, Chuan He, Xin Peng, Fan Lin, Chenxi Zhang, Shengfang Gong, Ziang Li,
  Jiayu Ou, and Zheshun Wu.
\newblock Microhecl: High-efficient root cause localization in large-scale
  microservice systems.
\newblock In {\em Proceedings of the 43rd International Conference on Software
  Engineering: Software Engineering in Practice}, ICSE-SEIP '21, page
  338–347. IEEE Press, 2021.

\bibitem{liu2019logzip}
Jinyang Liu, Jieming Zhu, Shilin He, Pinjia He, Zibin Zheng, and Michael~R Lyu.
\newblock Logzip: Extracting hidden structures via iterative clustering for log
  compression.
\newblock In {\em 2019 34th IEEE/ACM International Conference on Automated
  Software Engineering (ASE)}, pages 863--873. IEEE, 2019.

\bibitem{logzip}
Jinyang Liu, Jieming Zhu, Shilin He, Pinjia He, Zibin Zheng, and Michael~R.
  Lyu.
\newblock Logzip: extracting hidden structures via iterative clustering for log
  compression.
\newblock In {\em Proceedings of the 34th IEEE/ACM International Conference on
  Automated Software Engineering}, ASE '19, page 863–873. IEEE Press, 2020.

\bibitem{TraceAnomaly}
Ping Liu, Haowen Xu, Qianyu Ouyang, Rui Jiao, Zhekang Chen, Shenglin Zhang,
  Jiahai Yang, Linlin Mo, Jice Zeng, Wenman Xue, and Dan Pei.
\newblock Unsupervised detection of microservice trace anomalies through
  service-level deep bayesian networks.
\newblock In {\em 2020 IEEE 31st International Symposium on Software
  Reliability Engineering (ISSRE)}, pages 48--58, 2020.

\bibitem{pivot}
Jonathan Mace, Ryan Roelke, and Rodrigo Fonseca.
\newblock Pivot tracing: dynamic causal monitoring for distributed systems.
\newblock In {\em Proceedings of the 25th Symposium on Operating Systems
  Principles}, SOSP '15, page 378–393, New York, NY, USA, 2015. Association
  for Computing Machinery.

\bibitem{Meinig2019RoughLA}
Michael Meinig, Peter Tr{\"o}ger, and Christoph Meinel.
\newblock Rough logs: A data reduction approach for log files.
\newblock In {\em International Conference on Enterprise Information Systems},
  2019.

\bibitem{mi2013toward}
Haibo Mi, Huaimin Wang, Yangfan Zhou, Michael Rung-Tsong Lyu, and Hua Cai.
\newblock Toward fine-grained, unsupervised, scalable performance diagnosis for
  production cloud computing systems.
\newblock {\em IEEE Transactions on Parallel and Distributed Systems},
  24(6):1245--1255, 2013.

\bibitem{mongodb}
MongoDB.
\newblock Mongodb: The developer data platform | mongodb.
\newblock \url{https://www.mongodb.com/}, 2024.
\newblock Accessed Mar. 6, 2024.

\bibitem{mysql}
MySQL.
\newblock Mysql.
\newblock \url{https://www.mysql.com/}, 2024.
\newblock Accessed Mar. 6, 2024.

\bibitem{pinpoint}
Naver.
\newblock Pinpoint | leading open-source apm.
\newblock \url{ https://pinpoint-apm.gitbook.io/pinpoint/}, 2024.
\newblock Accessed: 2024/4/29.

\bibitem{tracead}
Sasho Nedelkoski, Jorge Cardoso, and Odej Kao.
\newblock Anomaly detection from system tracing data using multimodal deep
  learning.
\newblock In {\em 2019 IEEE 12th International Conference on Cloud Computing
  (CLOUD)}, pages 179--186, 2019.

\bibitem{opentelemetry}
Opentelemetry.
\newblock Opentelemetry.
\newblock \url{https://opentelemetry.io}, 2023.
\newblock Accessed: 2024/4/29.

\bibitem{ot-headsampling}
opentelemetry.
\newblock Opentelemetry head sampling.
\newblock \url{https://opentelemetry.io/docs/concepts/sampling/#head-sampling},
  2024.
\newblock Accessed: 2024/4/29.

\bibitem{instrumentation}
opentelemetry.
\newblock Opentelemetry instrumentation.
\newblock \url{https://opentelemetry.io/docs/concepts/instrumentation/}, 2024.
\newblock Accessed: 2024/4/29.

\bibitem{otlp}
opentelemetry.
\newblock Opentelemetry protocol.
\newblock https://opentelemetry.io/docs/specs/otel/protocol/, 2024.
\newblock Accessed: 2024/4/29.

\bibitem{span}
opentelemetry.
\newblock Opentelemetry span.
\newblock \url{https://opentelemetry.io/docs/concepts/signals/traces/#spans},
  2024.
\newblock Accessed: 2024/4/29.

\bibitem{ot-tailsampling}
opentelemetry.
\newblock Opentelemetry tail sampling.
\newblock \url{https://opentelemetry.io/docs/concepts/sampling/#tail-sampling},
  2024.
\newblock Accessed: 2024/4/29.

\bibitem{spectrum}
Thomas~W. Reps, Thomas Ball, Manuvir Das, and James~R. Larus.
\newblock The use of program profiling for software maintenance with
  applications to the year 2000 problem.
\newblock In {\em 6th European Software Engineering Conference Held Jointly
  with the 5th {ACM} {SIGSOFT} Symposium on Foundations of Software
  Engineering}, pages 432--449, 1997.

\bibitem{CLP}
Kirk Rodrigues, Yu~Luo, and Ding Yuan.
\newblock {CLP}: Efficient and scalable search on compressed text logs.
\newblock In {\em 15th {USENIX} Symposium on Operating Systems Design and
  Implementation ({OSDI} 21)}, pages 183--198. {USENIX} Association, July 2021.

\bibitem{Diagnosing}
Raja~R. Sambasivan, Alice~X. Zheng, Michael~De Rosa, Elie Krevat, Spencer
  Whitman, Michael Stroucken, William Wang, Lianghong Xu, and Gregory~R.
  Ganger.
\newblock Diagnosing performance changes by comparing request flows.
\newblock In {\em 8th USENIX Symposium on Networked Systems Design and
  Implementation (NSDI 11)}, Boston, MA, March 2011. USENIX Association.

\bibitem{sigelman2010dapper}
Benjamin~H Sigelman, Luiz~Andr{\'e} Barroso, Mike Burrows, Pat Stephenson,
  Manoj Plakal, Donald Beaver, Saul Jaspan, and Chandan Shanbhag.
\newblock Dapper, a large-scale distributed systems tracing infrastructure.
\newblock 2010.

\bibitem{bloom}
Sasu Tarkoma, Christian~Esteve Rothenberg, and Eemil Lagerspetz.
\newblock Theory and practice of bloom filters for distributed systems.
\newblock {\em IEEE Communications Surveys \& Tutorials}, 14(1):131--155, 2011.

\bibitem{loggrep}
Junyu Wei, Guangyan Zhang, Junchao Chen, Yang Wang, Weimin Zheng, Tingtao Sun,
  Jiesheng Wu, and Jiangwei Jiang.
\newblock Loggrep: Fast and cheap cloud log storage by exploiting both static
  and runtime patterns.
\newblock In {\em Proceedings of the Eighteenth European Conference on Computer
  Systems}, EuroSys '23, page 452–468, New York, NY, USA, 2023. Association
  for Computing Machinery.

\bibitem{logreducer21fast}
Junyu Wei, Guangyan Zhang, Yang Wang, Zhiwei Liu, Zhanyang Zhu, Junchao Chen,
  Tingtao Sun, and Qi~Zhou.
\newblock On the feasibility of parser-based log compression in
  $\{$Large-Scale$\}$ cloud systems.
\newblock In {\em 19th USENIX Conference on File and Storage Technologies (FAST
  21)}, pages 249--262, 2021.

\bibitem{improving}
Kundi Yao, Mohammed Sayagh, Weiyi Shang, and Ahmed~E. Hassan.
\newblock Improving state-of-the-art compression techniques for log management
  tools.
\newblock {\em IEEE Transactions on Software Engineering}, 48(8):2748--2760,
  2022.

\bibitem{Microrank}
Guangba Yu, Pengfei Chen, Hongyang Chen, Zijie Guan, Zicheng Huang, Linxiao
  Jing, Tianjun Weng, Xinmeng Sun, and Xiaoyun Li.
\newblock Microrank: End-to-end latency issue localization with extended
  spectrum analysis in microservice environments.
\newblock In {\em WWW 2021}, page 3087–3098. ACM, 2021.

\bibitem{yu2023logreducer}
Guangba Yu, Pengfei Chen, Pairui Li, Tianjun Weng, Haibing Zheng, Yuetang Deng,
  and Zibin Zheng.
\newblock Logreducer: Identify and reduce log hotspots in kernel on the fly.
\newblock In {\em 2023 IEEE/ACM 45th International Conference on Software
  Engineering (ICSE)}, pages 1763--1775. IEEE, 2023.

\bibitem{traceCRL}
Chenxi Zhang, Xin Peng, Tong Zhou, Chaofeng Sha, Zhenghui Yan, Yiru Chen, and
  Hong Yang.
\newblock Tracecrl: Contrastive representation learning for microservice trace
  analysis.
\newblock In {\em ESEC/FSE 2022}, page 1221–1232. ACM, 2022.

\bibitem{hindsight}
Lei Zhang, Zhiqiang Xie, Vaastav Anand, Ymir Vigfusson, and Jonathan Mace.
\newblock The benefit of hindsight: Tracing $\{$Edge-Cases$\}$ in distributed
  systems.
\newblock In {\em 20th USENIX Symposium on Networked Systems Design and
  Implementation (NSDI 23)}, pages 321--339, 2023.

\end{thebibliography}
